\documentclass{aa}  
\usepackage{url}
\usepackage{txfonts}
\usepackage{xspace}
\usepackage{natbib}
\usepackage{caption}
\usepackage{float}
\bibpunct{(}{)}{;}{a}{}{,}
\def\gaia{\textit{Gaia}\xspace}
\def\kepler{\textit{Kepler}\xspace}
\usepackage[colorlinks=true, linkcolor=blue, citecolor=blue, urlcolor=blue]{hyperref}

\begin{document}

   \title{Automated all-sky detection of $\gamma$ Doradus / $\delta$ Scuti hybrids in TESS data from positive unlabelled (PU) learning}

   \author{Mykyta~Kliapets
          \inst{1,}\inst{2}
          \and
          Pablo~Huijse
          \inst{1}
          \and
          Andrew~Tkachenko
          \inst{1}
          \and
          Alex~Kemp
          \inst{1}
          \and
          Dario~J.~Fritzewski
          \inst{1}
          \and
          Daniel~Hey
          \inst{3}
          \and
          Conny~Aerts
          \inst{1,}\inst{4,}\inst{5}
          }

   \institute{Institute of Astronomy, KU Leuven, Celestijnenlaan 200D, bus 2401, 3001 Leuven, Belgium\\
              \email{mykyta.kliapets@kuleuven.be}
         \and
             Kavli Scholar funded by The Kavli Foundation, 5715 Mesmer Avenue, Los Angeles, CA 90230, USA
         \and
             Institute for Astronomy, University of Hawai’i, Honolulu, HI 96822, USA
        \and
            Department of Astrophysics, IMAPP, Radboud University Nijmegen, PO Box 9010, 6500 GL, Nijmegen, The Netherlands
        \and
            Max Planck Institute for Astronomy, Koenigstuhl 17, 69117, Heidelberg, Germany
             }

   \date{Received TBA; accepted TBA}
 
  \abstract
   {The Transiting Exoplanet Survey Satellite (TESS) mission has observed hundreds of millions of stars, substantially contributing to the available pool of high-precision photometric space data. Among them are the relatively rare $\gamma$ Doradus / $\delta$ Scuti ($\gamma$ Dor / $\delta$ Sct) hybrid pulsators, which have been previously studied using \kepler data. These stars are perfect laboratories to probe both inner and outer interior stellar layers thanks to them exhibiting both pressure and gravity modes.}
   {We seek to classify an all-sky sample of AF stars observed by TESS to find previously undiscovered hybrid pulsators and supply them in a catalogue of candidates. We also aim to compare the light curves produced with the TESS-\gaia Light Curve (TGLC) pipeline, currently underused in variability studies, with other publicly available light curves.}
   {We compared dominant and secondary frequencies of confirmed hybrid pulsators in \kepler, extended mission Quick Look Pipeline (QLP) data, and nominal and extended mission TGLC data. We then used a feature-based positive unlabelled (PU) learning classifier to search for new hybrid pulsators amongst TESS AF stars and investigated the properties of the detected populations.}
   {We find that the variability of confirmed hybrids in TGLC agrees well with the one occurring in QLP light curves and has a high recovery rate of \kepler-extracted frequencies. Our `smart binning' method allows for robust extraction of hybrids from large unlabelled datasets, with an average out-of-bag prediction for test set hybrids at 93.04\%. The analysis of dominant frequencies in high-probability candidates shows that we find more pressure-mode dominant hybrids. Our catalogue includes 62,026 new candidate light curves from the nominal and extended TESS missions, with individual probabilities of being a hybrid in each available sector.}
   {Our catalogue results in a major increase of TESS $\gamma$ Dor / $\delta$ Sct hybrid pulsators, suitable for further asteroseismic studies.}

   \authorrunning{M. Kliapets et al.} 
   \titlerunning{PU Learning of Hybrid $\gamma$ Dor / $\delta$ Sct Pulsators in TESS}
   \maketitle
%
%-------------------------------------------------------------------

\section{Introduction}
   Stellar variability holds a special place in the theory of stellar structure and evolution \citep{eyer2023}. Changes in brightness over time reveal a wealth of information about the inner workings of stars themselves, as well as their interactions with companions such as other stars and exoplanets. Over the last several decades, the field of asteroseismology \citep{aerts2010} has significantly benefited from nearly continuous high-precision photometry available from space missions. These space telescopes take advantage of the absence of atmospheric effects, absorption of certain parts of the electromagnetic spectrum, day-night cycles, and negative effects of light pollution. This results in a significant increase in the quality and quantity of continuous data available to the variable star community.
   
   Variable star studies have been revolutionized by the high-precision long-baseline data of more than 160,000 stars provided by the \kepler \citep{borucki2010} satellite, including the K2 mission \citep{howell2014}, as well as ongoing observations of hundreds of millions of stars by TESS \citep{ricker2015}. Furthermore, \gaia \citep{brown2016} has also significantly contributed to asteroseismic research \citep{deridder2023}, despite not being explicitly designed for variable star studies. Many additional discoveries are expected with the upcoming ESA PLAnetary Transits and Oscillations (PLATO, \citeauthor{rauer2025} \citeyear{rauer2025}) mission, particularly in the context of its complementary science programme, which is set to explore all science cases not connected to the asteroseismology of exoplanet-hosting main-sequence stars. 
   
   Within the myriad of targets observed from space, OBAF pulsators — both radial and non-radial — are of particular interest to asteroseismologists, enabling precise testing of stellar evolution theories across a wide mass regime \citep{deridder2023}. Photometric data with $\mu$mag precision has enabled the study of these objects in various domains, including the investigation of transport processes (e.g. \citeauthor{mathis2009} \citeyear{mathis2009}, \citeauthor{rogers2015} \citeyear{rogers2015}, \citeauthor{eggenberger2019} \citeyear{eggenberger2019}), calculations of convective core masses (e.g. \citeauthor{mombarg2024} \citeyear{mombarg2024}),  near-core rotation frequency estimations (e.g. \citeauthor{aerts2025} \citeyear{aerts2025}), magnetic field studies (e.g. \citeauthor{mathis2021} \citeyear{mathis2021}, \citeauthor{li2022} \citeyear{li2022}), multiplicity (e.g. \citeauthor{ijspeert2021} \citeyear{ijspeert2021}), and many others. For the most complete summary of recent findings, we refer the reader to the reviews by \cite{kurtz2022} and \cite{aerts2024}.

   The stellar interior is probed by two types of stellar oscillations. In main-sequence stars, pressure (p) modes probe outer stellar regions and have frequencies higher than approximately 4 d$^{-1}$, while gravity (g) modes are sensitive to near-core regions and have frequencies below approximately 4 d$^{-1}$ \citep{chapellier2012}. However, the instability regions of the Hertzsprung–Russell diagram can overlap where stars exhibit both p and g~modes (see Fig. 1, \citeauthor{aerts2021} \citeyear{aerts2021}). Here, we study one such case of overlapping regimes, particularly $\gamma$ Dor / $\delta$ Sct hybrids. These hybrid pulsating stars (\citeauthor{dupret2005} \citeyear{dupret2005}, \citeauthor{grigahcene2010} \citeyear{grigahcene2010}) are especially interesting for asteroseismology, as they allow us to more precisely constrain stellar structure and evolution models of stars slightly more massive than the Sun — which have a convective core — thanks to the ability to probe outer and inner stellar interior layers simultaneously \citep{bradley2015}. It is thus not surprising that this (sub-)class of pulsating stars received attention in both observational asteroseismology (e.g. \citeauthor{bradley2015} \citeyear{bradley2015}) and asteroseismic modelling (e.g. \citeauthor{zhang2020} \citeyear{zhang2020}).
   
   Studies of $\gamma$ Dor / $\delta$ Sct hybrids are limited by the scarcity of objects readily available for analysis. Hundreds of stars previously believed to be pure $\gamma$ Dor and $\delta$ Sct pulsators observed by \kepler (\citeauthor{bowman2016} \citeyear{bowman2016}, \citeauthor{li2020} \citeyear{li2020}) were later revealed to be hybrids (e.g. \citeauthor{audenaert2022} \citeyear{audenaert2022}). Despite that, many have yet to be discovered in the TESS data (as demonstrated by \citeauthor{skarka2022} \citeyear{skarka2022} and \citeauthor{skarka2024} \citeyear{skarka2024}), which are orders of magnitude more voluminous in the number of targets observed compared to \kepler. Manual inspection of millions of light curves of (AF) stars is impractical and thus requires automated processing algorithms to assist with the selection of candidates for further asteroseismic studies.
   
   Asteroseismology has increasingly relied on machine learning when working with large observational datasets. A wide variety of machine learning frameworks have been used in variability studies, with some aimed at $\gamma$ Dor / $\delta$ Sct hybrids specifically \citep{audenaert2022}. Machine learning allows one to take population studies and ensemble asteroseismology to novel heights, enabling efficient data handling and proposing candidates for more detailed follow-ups. For a summary of the current state of machine learning in asteroseismology, we refer the reader to the review by \citet{audenaert2025}. 

   Supervised machine learning, used when a dataset of labelled by experts is available, has seen extensive use in variable star studies. For example, \citet{audenaert2021} used an ensemble of supervised learners, including a meta-classifier combining the output of other classifiers, on a labelled dataset with eight classes. \citet{hey2024} used a random forest classifier on a dataset of six classes of variable stars, including hybrid pulsators. Some other applications include looking for a specific type of variable stars \citep{van2018} and multi-survey classifications \citep{aguirre2019}.

   Unsupervised machine learning, used to discover hidden patterns in the data instead of assigning objects to particular classes, has also seen use in asteroseismology. \citet{audenaert2022} used an unsupervised clustering algorithm to discern populations of pulsators in a dataset of $\gamma$ Dor and $\delta$ Sct stars. \citet{rizhko2024} used a self-supervised multi-modal contrastive learning to study a dataset of variable stars. Interestingly, \gaia data has been subject to a number of unsupervised machine learning applications, both within the domain of variability studies and beyond. For example, \citet{huijse2025} performed an all-sky variability analysis on the entire \gaia DR3 dataset. \citet{ranaivomanana2025} used an unsupervised clustering algorithm to detect sub-populations of hot sub-luminous stars. One undisputed advantage of unsupervised algorithms is that they are free of human bias, meaning they are not constrained by the pre-defined labels as in supervised learning, which makes them well-suited to detect novel (sub-)classes of variable stars and check scientific intuition.

   While the output of supervised machine learning is easier to evaluate than unsupervised learning, its main advantage — reliance on a large labelled set — is also its disadvantage when it comes to variability studies of rare classes such as $\gamma$ Dor / $\delta$ Sct hybrids. Finding more examples from a rare class of astrophysical objects requires having a large set of labelled examples. However, those labels would have to come from manually inspecting the unlabelled data, a process we have already shown to be impractical. This `vicious circle' slows down the progress of variability classification of rare pulsators, and subsequently their further asteroseismic follow-up studies. There are at least three ways to tackle this problem.

    One potential solution is to use algorithms specifically designed to handle class imbalance, known as imbalance learning \citep{chen2024}. \citet{hosenie2020}, \citet{kang2023}, and \citet{zhang2023} have used various imbalance learning architectures to classify periodic variables, allowing for a better classification performance in situations where objects of one class could outnumber objects of other classes. Another option to consider is a family of anomaly detection algorithms. This paradigm deals with the detection of data points that are significantly different from the majority of data (`normal' observations). These algorithms can be used to both find novel astrophysical phenomena (e.g. \citeauthor{rebbapragada2009} \citeyear{rebbapragada2009}) and look for specific rare objects (e.g. \citeauthor{sanchez2021} \citeyear{sanchez2021} and \citeauthor{chan2022} \citeyear{chan2022}). Interestingly, this paradigm comprises both supervised and unsupervised anomaly detection algorithms, in contrast to imbalance learning algorithms that require labelled instances. For example, \citet{huijse2025} have used unsupervised outlier detection methods for error analysis of their unsupervised clustering framework. Unsupervised anomaly detection, however, does not have any prior knowledge on anomalies \citep{pang2021}, so the objects singled out by the algorithm might not belong to the class of interest in the first place, while supervised anomaly detection also relies on time-intensive labelling of both `normal' and anomalous observations \citep{pang2021}.

   Finally, it is possible to use semi-supervised learning by taking advantage of the unlabelled examples to learn the structure and to regularize the classification boundaries learned from the scarce labelled examples \citep{vanengelen2020}. These algorithms remedy situations where labelled data is difficult to obtain, such as in medical imaging and risk assessment applications. This circumvents the need to label most of the dataset, and instead label only some of the positive (class of interest) and/or negative (other) data. Although this approach might appear ideal, semi-supervised machine learning shares evaluation and interpretation challenges of unsupervised learning, which might explain why it has not seen a lot of use in variability studies so far. In this work, we aim to rectify this research gap and demonstrate the potential of semi-supervised learning for rare object detection in asteroseismology.

   The most common object of study in variability classification is a light curve. Most data-driven applications for asteroseismology relying on TESS data have so far not used the light curves produced with TGLC \citep{han2023} pipeline. Until now, the focus has been on other science products available from the Mikulski Archive for Space Telescopes (MAST),\footnote{https://archive.stsci.edu} namely QLP (\citeauthor{huang2020a} \citeyear{huang2020a}, \citeyear{huang2020b}, \citeauthor{kunimoto2021} \citeyear{kunimoto2021}, \citeyear{kunimoto2022}) and TASOC (\citeauthor{handberg2021} \citeyear{handberg2021}, \citeauthor{lund2021} \citeyear{lund2021}) light curves. Our work therefore, has two connected goals:

   \begin{itemize}
       \item To compare the dominant and secondary variability of confirmed hybrid pulsators in the TGLC data from the nominal (sectors 1-26, 30-minute cadence data) and extended (sectors 27-, 10-minute cadence data) TESS mission with other available light curves (QLP and \kepler data); and

       \item To use a semi-supervised machine learning method on a dataset of known hybrids and unlabelled set of TESS observations to create a large all-sky catalogue of candidate hybrid pulsators with individual probabilities of being a $\gamma$ Dor / $\delta$ Sct hybrid in a single sector and multi-mode variability extracted from TGLC light curves. Our new public catalogue can then be used by the asteroseismology community for follow-up studies, including unsupervised machine learning clustering, population studies, mode identification, asteroseismic modelling, and probing for other astrophysical properties.
   \end{itemize}

%--------------------------------------------------------------------

\section{TGLC for hybrid pulsator studies}

   Several high-level science products (HLSP) based on raw TESS data are available. In this Section, we describe the TGLC \citep{han2023} pipeline and compare the available data for known hybrid pulsators with other publicly available data, particularly \kepler and QLP light curves.

\subsection{The TGLC pipeline}

    A complete description of the TGLC methodology has been given by \citet{han2023}, so here we only briefly summarize it for convenience. The TGLC pipeline allows to readily access light curves up to 16th magnitude by taking advantage of \gaia's astrometry and photometry, particularly of the high angular resolution offered by \gaia \citep{brown2016}. This is in contrast to other available pipelines, where only sources up to 13.5 magnitude are readily available for download on MAST.

\begin{figure}
    \centering
    \resizebox{\hsize}{!}{\includegraphics{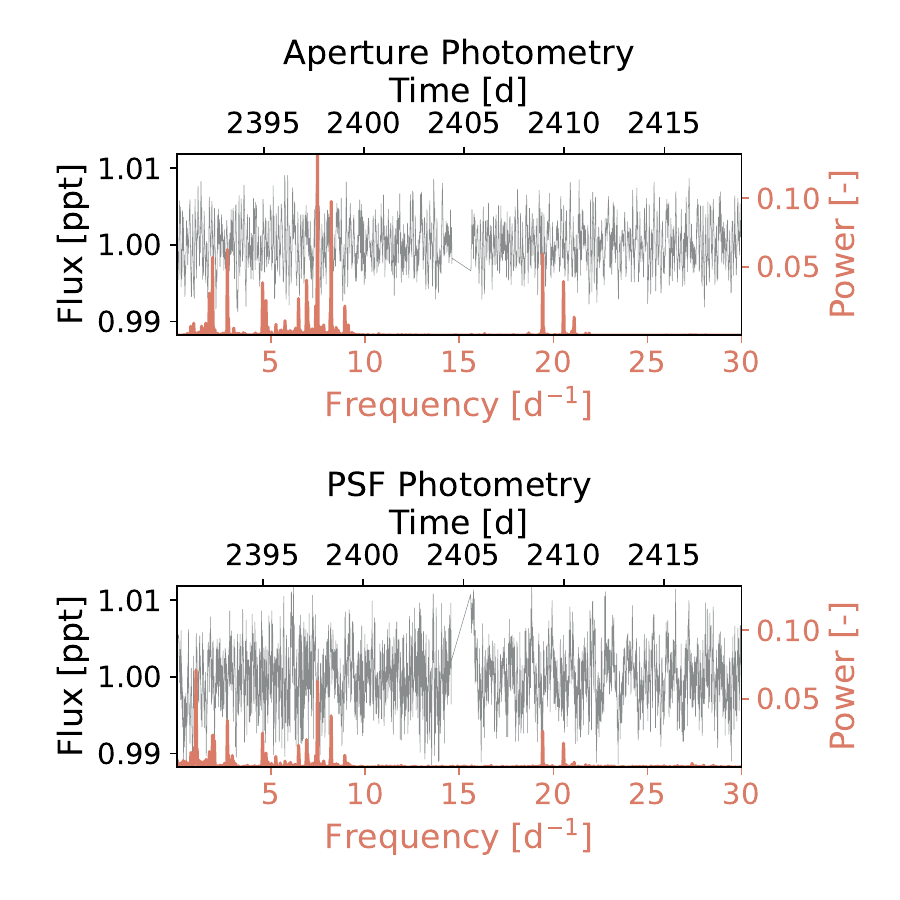}}
    \caption{Aperture (top) and PSF (bottom) light curve (grey) and Lomb-Scargle periodogram (overplotted in burgundy, clipped at 30 d$^{-1}$ for visibility) of a confirmed hybrid pulsator (DR3 2147267632621883776) detected in TESS sector 40. Note how this target is p- and g-mode dominated on the aperture and PSF light curves, respectively.}
    \label{fig:ap_psf}
\end{figure}

    TGLC provides both point spread function (PSF) and aperture light curves for each source. Both rely on the position and brightness of stars in the field from \gaia to construct the final science product. In general, an aperture TGLC light curve delivers higher-quality data than a PSF one in crowded fields. This leads to more reliable amplitude estimations for variable stars \citep{han2023}. In Fig.~\ref{fig:ap_psf} we show an example of both a PSF and an aperture light curve of a confirmed hybrid pulsator with their respective Lomb-Scargle periodograms. For that target, we find a similar peak structure for both periodograms, although the one for aperture photometry exhibits higher amplitudes. The standard deviation of the flux for the PSF light curve is twice that of the aperture one. This is caused by the noisier floor throughout the entire frequency space. 
    
    We additionally manually inspected light curves of hybrids and a random subsample of other sources in TGLC (see Section \ref{sec:data}). We find no pathological long-term instrumental trends in light curves in both science products. However, we note an occasional systematic at 1 d$^{-1}$, previously reported by \citet{hey2024}, which appears in PSF light curves more often.  We further found that calibrated PSF flux is missing for 5.4\% of our sample. We therefore opted to use aperture light curves, both to have the most reliable amplitude estimations for variable stars and for completeness of the dataset. The remainder of this Section focuses on comparing the two dominant independent frequencies ($f_{1}$ and $f_{2}$) extracted from the aperture TGLC light curves with other light curves available on MAST for the confirmed hybrids. 

\subsection{Dominant and secondary variability of labelled hybrids}\label{sec:var}
    
    We preprocessed each aperture TGLC light curve using both the normal TESS\footnote{https://archive.stsci.edu/files/live/sites/mast/files/home/missions-and-data/active-missions/tess/\_documents/EXP-TESS-ARC-ICD-TM-0014-Rev-F.pdf} and TGLC quality flags \citep{han2023}, clipping flux outliers above 3 standard deviations, and applying a high-pass Gaussian filter to remove long-term trends caused by the thermomechanical behaviour of the satellite, which does not contain any astrophysical information. More concretely, we subtracted the Gaussian smoothed time series with $\sigma_{G}=100$ datapoints to remove long-period instrumental trends. For hybrids from the extended TESS mission, we found that the ratio of the standard deviation of the subtracted signal to the original ($\sigma_{subtr}/\sigma_{original}$) is $0.107 \pm 0.066$, suggesting that the preprocessing pipeline removes signal comprising one tenth of the entire light curve variability, which could be instrumental. For a subset of non-hybrid sources, the ratio of $\sigma_{subtr}/\sigma_{original}$ was expectedly higher at $0.199 \pm 0.114$, as signals in a largely non-variable set are dominated by instrumental trends, noise, and systematics removed by preprocessing, rather than genuine (intrinsic) variability that passes the filter.

\subsubsection{Effect of \gaia-informed light curve extraction: TGLC vs. QLP light curves}

    Differences in light curve extraction pipelines can have a significant impact on the light curve, as well as the frequency content detected in the periodogram, and the amplitudes of pulsations (\citeauthor{han2023} \citeyear{han2023}, \citeauthor{hubrig2024} \citeyear{hubrig2024}, \citeauthor{kunimoto2024} \citeyear{kunimoto2024}). The QLP light curves remain one of the most used in variability studies and offer greater sector availability on MAST compared to TGLC. We therefore compared the dominant and secondary variability for the hybrid pulsators available in both pipelines by producing frequency-frequency plot from preprocessed data, shown on Fig.~\ref{fig:A1}. We applied the same preprocessing to both types of light curves for a fair comparison (see Section \ref{sec:var}). Both light curves agree well for both extracted frequencies, having an agreement of 77.13\% and 56.44\% within 0.1 d$^{-1}$ of the unity line, respectively. There is noticeable scatter for $f_{2}$ (bottom) both above twice unity and below half-unity, where one pipeline detects a high frequency and another a low frequency, or vice versa. This suggests that modes with low amplitudes are susceptible to differences in processing pipelines and noise. The agreement is generally better at higher frequencies, partially caused by the occasional TGLC systematic at 1 d$^{-1}$, which is extracted as a dominant peak, while a genuine mode might be recovered by the QLP.

\subsubsection{Longer baselines for g-mode detection: TESS TGLC and QLP vs. \kepler light curves}\label{sec:kepler}

\begin{figure}[htbp]
    \centering
    \resizebox{\hsize}{!}{\includegraphics{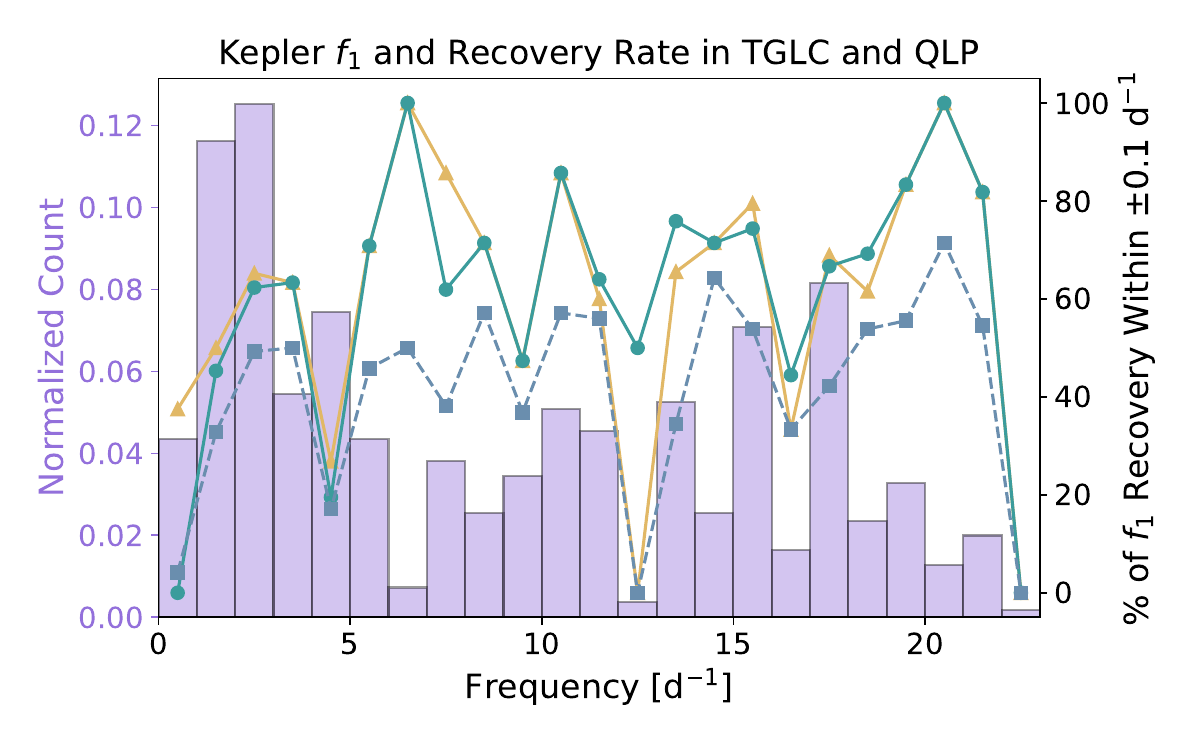}}
    \caption{Comparison of \kepler $f_{1}$ (purple histogram) recovery efficiency in QLP (gold), TGLC (genuine and binned to 30-minute cadence in teal and grey, respectively).}
    \label{fig:recovery}
\end{figure}

    Single-sector TESS observations are typically too short to correctly detect all g~modes. The four-year \kepler data, on the other hand, has proven to be a treasure trove for the studies of g-mode oscillators (\citeauthor{uytterhoeven2011} \citeyear{uytterhoeven2011}, \citeauthor{tkachenko2014} \citeyear{tkachenko2014}, \citeauthor{li2020} \citeyear{li2020}, \citeauthor{pedersen2021} \citeyear{pedersen2021}, \citeauthor{aerts2024} \citeyear{aerts2024}). We therefore investigated the efficiency of recovering \kepler-extracted frequencies. Fig.~\ref{fig:recovery} shows that the frequency recovery rate for $f_{1}$ is approximately the same in both TGLC and QLP, although the systematic at 1 d$^{-1}$ affects the recovery rate of the lower frequencies. We found 50.88\% and 43.42\% recovery rates for QLP and TGLC, respectively. At higher frequencies, the two pipelines agree better, within <1\% differences in measured frequencies, allowing us to conclude that both pipelines are of similar quality for hybrid pulsator studies, noting the potential to find faint hybrids with TGLC light curves. While longer-baseline observations would lead to higher performance of any automated classification pipeline, concatenating several TESS sectors together causes issues of including data with inconsistent quality. We therefore opted to work with the shorter-baseline individual TESS sector data.

\subsubsection{Finer sampling for p-mode detection: primary- vs extended-mission TGLC light curves}

    The detection of p~modes is known to benefit from data with shorter cadence \citep{murphy2012}. We compared the dominant and secondary variability of confirmed hybrids in the extended and nominal TESS missions. We also considered a binned version of the extended mission where every three observations are averaged to simulate the 30-minute cadence. As seen in the middle and left panels of Fig.~\ref{fig:A1}, the agreement between the extended and nominal mission data for $f_{1}$ and $f_{2}$ is 74.33\% and 44.68\%, respectively. Between the 10- and 30-minute cadence extended mission data, the agreement is 90.21\% and 67.46\% within 0.1 d$^{-1}$ of the bisector for dominant and secondary variability, respectively. In both cases, more p~modes are detected in extended mission TGLC indicated by the scatter below the half-unity line. The quality of the nominal mission data is lower than the downsampled (binned) extended mission. The better result for 10- versus 30-minute cadence extended mission data comparison can be explained by them coming the same sector, whereas comparisons with nominal mission data are based on different sectors, capturing systematics in some sector data but not the other. 
    
    The binned TGLC data also sees a significant drop in \kepler recovery rate, where the detection of p~modes is affected more than g~modes (Fig.~\ref{fig:recovery}). The secondary frequency $f_{2}$ is more sensitive to the lower frequency resolution because of its lower amplitude (Fig.~\ref{fig:A1}). While the performance of 30-minute cadence data is undoubtedly lower than for 10-minute cadence data, we argue that it is justified to use both in separate training sets, provided their sufficient purity. Manual or semi-automatic vetting of light curves, ensuring that the hybrid pulsation signal is the one captured in the light curve rather than a systematic in the training set, should allow new candidates to be detected. We also note that sampling differences between the extended and nominal mission warrant separate classifications when frequency- and/or period-domain features are used, as the higher Nyquist limit would allow one to capture more p-mode variability in extended mission data. In practice, the median Nyquist frequency of the labelled data we used from nominal and extended missions amount to $24.00020 \pm 0.00011$ and $72.0009 \pm 0.0018$\,d$^{-1}$, respectively.

%--------------------------------------------------------------------
\section{PU learning}

    The simplest classification settings in machine learning is binary classification. In a binary classifier, a model is tasked with assigning instances of a tuple $(x,y)$, where $x$ are the features and $y\in\{0,1\}$ is a label, to one of two classes: positive $(y=1)$ or negative $(y=0)$ \citep{muller2016}. Being a supervised learning method, the classifier must be provided with labelled examples of both positive and negative instances. In the case of rare objects such as $\gamma$ Dor / $\delta$ Sct hybrids, providing a large enough labelled set would require a substantial manual effort, particularly to label a representative set of negative instances. 

    In Positive Unlabelled (PU) Learning, a family of semi-supervised algorithms, only part of the data is labelled. Instances in PU Learning are instead represented by a tuple $(x,y,s)$, where $s\in\{0,1\}$ denotes whether an instance was selected to be labelled (\citeauthor{jaskie2019} \citeyear{jaskie2019}, \citeauthor{bekker2020} \citeyear{bekker2020}). An important differentiating feature of PU Learning is that only positive instances are labelled, but not all of them. This means that an unobserved instance $(s=0)$ can be either positive $(y=1)$ or negative $(y=0)$, while observed instances $(s=1)$ are always positive: $p(y=1|s=1)=1$. An alternative way to reason about this is that the probability of a negative instance being labelled should be zero: $p(y=0|x,s=1)=0$ \citep{elkan2008}, unless it has been mistakenly labelled as positive by an expert. A selection of labelled instances is guided by one of the three labelling mechanisms, so a labelled distribution is a biased distribution of all positive instances: Selected Completely at Random (SCAR), Selected at Random (SAR), or Probabilistic Gap (a special case of SAR where positive instances are less likely to be labelled if they are closer to the negative ones in their attributes). Additionally, both positive (labelled) and unlabelled datasets are assumed to be independently and identically distributed (i.i.d.) samples of their respective real distributions. The data are assumed to be separable, meaning there is a classifier that can separate positive instances \citep{bekker2020}. A schematic representation of such a learning setting is shown in Fig.~\ref{fig:pu}. Several algorithms exist in the PU Learning family, from simpler two-step methods training binary classifiers to more complex probabilistic approaches and deep PU Learning (\citeauthor{jaskie2019} \citeyear{jaskie2019}, \citeauthor{li2022pu} \citeyear{li2022pu}, \citeauthor{bekker2020} \citeyear{bekker2020}). For the most complete summary, we refer the reader to the review by \citet{bekker2020}.
    
    Despite its advantage when searching for objects with few labels in large datasets \citep{yang2021}, PU Learning has not yet seen much use in astrophysics. One notable exception is the search for Carbon stars in the Sloan Digital Sky Survey \citep{du2016}. A similar learning approach, One-Class Classification, has been applied to look for quasars in Wide-Field Infrared Survey Explorer data \citep{solarz2017}. In general, however, PU Learning and semi-supervised learning in general remain rather underutilized in variability studies, likely caused by the evaluation and validation complexity.

%--------------------------------------------------------------------
\section{Application of PU learning to TESS data}

    In this Section, we outline the four key stages of our machine learning pipeline: data acquisition and preprocessing, feature engineering and extraction, model selection and training, and performance evaluation. As PU Learning requires having both a positive (labelled) and a (typically larger) unlabelled dataset, we begin by describing the two below.

\subsection{The dataset}\label{sec:data}

\subsubsection{Positive (labelled) dataset of confirmed hybrids}
    Our initial labelled dataset included published hybrids detected in the northern continuous viewing zone (CVZ) of TESS \citep{skarka2022} and in the \kepler field of view (\citeauthor{bowman2016} \citeyear{bowman2016}, \citeauthor{li2020} \citeyear{li2020}). We used two types of TESS light curves for these objects. First, we used the light curves available for download on MAST, which include 3,292 and 785 light curves from the nominal and extended missions, respectively. Second, because there are less TESS light curves available for the extended mission classification, we used the TGLC code made available by \citet{han2023} to extract TESS light curves from Full Frame Images for the extended mission.

    The confirmed hybrids from \citet{skarka2022} indicate the dominant variability of the target ($\gamma$ Dor / $\delta$ Sct or $\delta$ Sct /  $\gamma$ Dor hybrids). For the purposes of this study we do not differentiate between the two, simply labelling all datasets as hybrids. The dominant variability of hybrids (Fig.~\ref{fig:A1}) across the entire labelled set reveals that there are more p-mode dominated hybrids in the training set. This is expected to propagate into the classification (see Section \ref{sec:catalog}) by favouring $\delta$ Sct / $\gamma$ Dor hybrids from the unlabelled set. We suggest that this is partly due to the short baseline of TESS, which makes the detection of g~modes difficult.

    Additionally, not every source labelled as hybrid in one survey will manifest as a hybrid in another. Since part of the labelled sample comes from \kepler, some of the g~modes will not be resolved in TESS light curves due to different observational baselines. Removing these sources is a necessary step to avoid confusing the classifier and assigning high probabilities to other (non-hybrid) light curves, which defeats the purpose of an automated detection pipeline. We therefore manually vetted each TGLC light curve and their respective periodograms to ensure that only high-quality observations with low noise levels and systematics are used for training. After manual inspection of both publicly available and manually generated light curves, our labelled dataset comprises 480 unique sources, totalling 1,404 light curves in the nominal and 2,376 light curves in the extended mission.

\subsubsection{Unlabelled dataset of AF stars in TESS}

    The unlabelled dataset comprises the aperture TGLC data for stars that are available to download on MAST (up to TESS sector 44 as of 31 March 2025). We used the same filtering approach to work with the TESS Input Catalog as described in \citet{ijspeert2021}, except that our cut is based on the TIC effective temperature ($T_{\mathrm{eff}}$) instead of colours. We limited our selection to stars with $10,000 \geq T_{\mathrm{eff}} \geq 4,000$ K, where the lower limit is generously placed below the temperature limit of solar-like oscillators to account for potential errors in measured temperatures. The upper limit rejects most Slowly Pulsating B-type Stars / $\beta$ Cep (SPB / $\beta$ Cep) hybrids and p- and g-mode pulsating subdwarfs (sdBVs), which are outside the scope of this work.

    Applying the $T_{\mathrm{eff}}$ cut yielded 29,309,213 light curves for dwarf stars, 16,168,167 in the nominal mission and 13,141,046 in the extended mission. The number of light curves differs significantly from sector to sector, with the smallest set coming from sector 4 (49,759) and the largest from sector 12 (1,716,982). Instances in the positive (labelled) set were removed from the unlabelled set before training regardless of the sector to avoid rediscovering hybrids that have already been found.

    Application of a PU Learning algorithm to this dataset is based on an assumption that there are instances in this unlabelled dataset that belong to the positive class of hybrid pulsators, which have not been labelled (discovered) yet (Fig.~\ref{fig:pu}). In our case, it is based on the lack of all-sky hybrid searches in the TESS data, as previous studies have been limited to the northern \citep{skarka2022} or southern \citep{skarka2024} CVZ of TESS, or a limited number of TESS sectors \citep{audenaert2022}.

\subsection{Classification features}\label{sec:features}

\begin{figure}
    \centering
    \resizebox{\hsize}{!}{\includegraphics{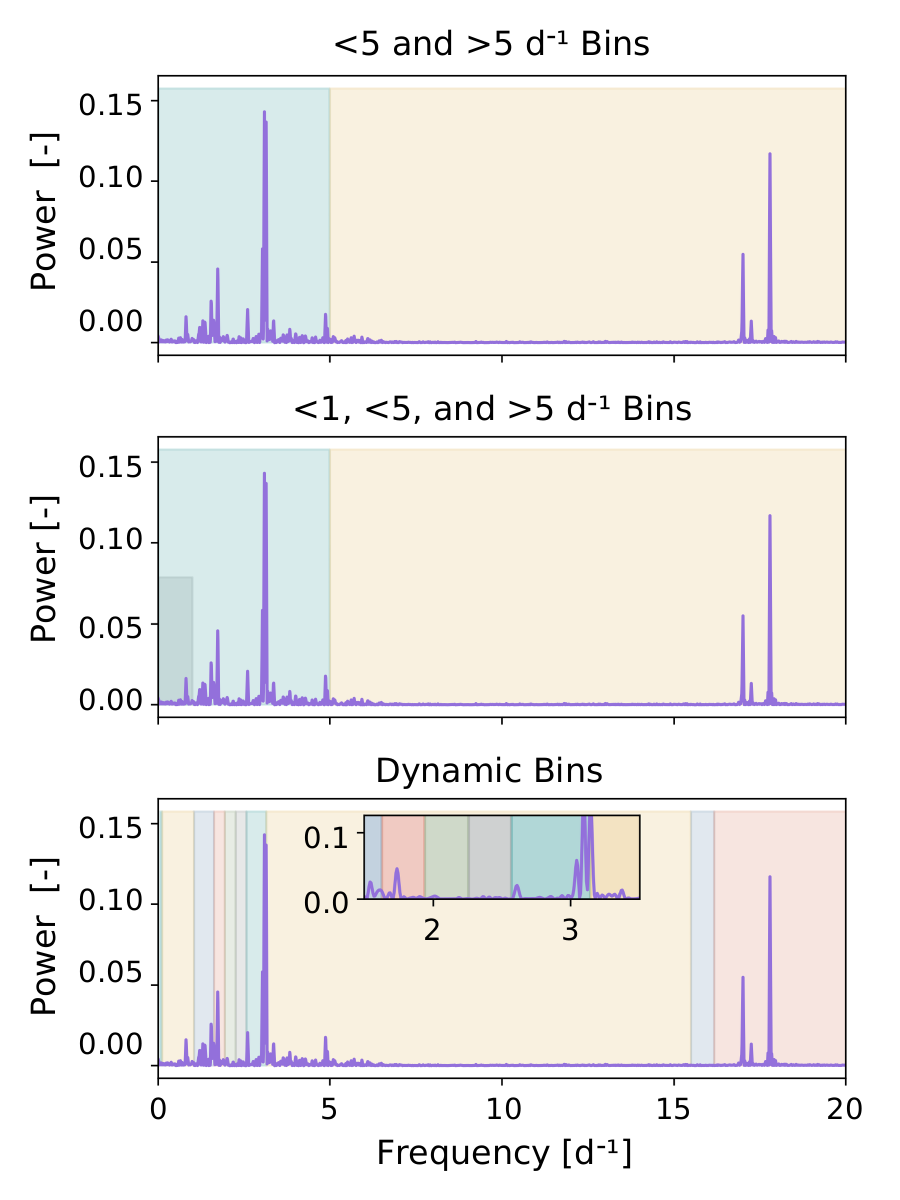}}
    \caption{Graphic representation of binning approaches applied to a confirmed hybrid pulsator (DR3 1653802003712212992). Top plot: bins from \citet{skarka2022} and \citet{skarka2024}, centre plot: bins from \citet{hey2024}, bottom plot: this work.}
    \label{fig:bins}
\end{figure}

    The search for hybrid pulsators, whether manual or automatic, relies on variability analysis, often resulting in the use of frequency-domain features, such as frequencies or periods, amplitudes, and phases (\citeauthor{debosscher2007} \citeyear{debosscher2007}, \citeauthor{vanderplas2018} \citeyear{vanderplas2018}). As hybrid pulsators exhibit variability in two different frequency regimes, previous studies used binning to extract this information, which involves dividing the frequency space into smaller parts, from which features are extracted. There are several approaches to periodogram binning: (i) bins can either be equal or vary in size; and (ii) bins can be overlapping or non-overlapping. \citet{skarka2022} and \citet{skarka2024} previously used two bins to look for $\gamma$ Dor / $\delta$ Sct hybrids within a population of AF stars: $<5\,\mathrm{d}^{-1}$ and $>5\,\mathrm{d}^{-1}$. \citet{hey2024} used three partially overlapping bins in their analysis of a population of OBAF pulsators: $<1\,\mathrm{d}^{-1}$, $<5\,\mathrm{d}^{-1}$, and $>5\,\mathrm{d}^{-1}$. A graphic representation of these two approaches is shown on Fig.~\ref{fig:bins} (top and middle).

    In this work, we used a dynamic binning approach to automatically find the most optimal bins for hybrid detection. For each labelled hybrid, we preprocessed the light curve, computed a frequency spectrum using a Lomb-Scargle periodogram method \citep{vanderplas2018} with $f_{min}=10^{-4}$ d$^{-1}$ and $f_{max}=\frac{1}{2\Delta t}$. This low lower limit is inspired by the fact that pulsators with low-frequency g~modes have beating periods that may span many years, notably longer than the 4-year nominal \kepler light curves \citep{VanBeeck2021}. 
    We then identified peaks that are the local optima with \texttt{scipy.signal.findpeaks} and created a power histogram with $m$ bins. From the histogram, we calculated the cumulative power distribution across the $m$ frequency bins, normalized it, and applied $k$ power-law scaling. We then applied linear interpolation of $n$ linearly spaced bins to map them on the cumulative power distribution. This resulted in $n$ non-overlapping non-equal bins, which are finer in more `important' parts of the frequency space (where more peaks are present) and sparser bins elsewhere. We repeated this process for all labelled hybrids, averaging the bin edges for the entire labelled dataset, separate for nominal and extended missions hybrids. A graphic representation of our approach with $m=50$, $n=10$, and $k=1$ (no power law scaling) is shown on Fig.~\ref{fig:bins} (bottom). The averaged bins were then applied to the entire unlabelled set to extract features for classification, ensuring that the bins are best-fit to look for a specific type of pulsators. From each of the bins we extracted the amplitude of the highest peak and the four statistical moments: mean, variance, skewness, and kurtosis, with higher feature values in multiple relevant bins meant to separate hybrids from pure p- or g-mode pulsators. The same was repeated for the period space, resulting in two sets of features for each light curve. 

    There are two reasons why we only extracted the amplitudes from the bins and not the highest frequencies themselves. First, the bin widths themselves serve as proxies for frequencies and periods, especially in the finer-spaced bins. Amplitudes therefore allow for the extraction of additional information about variability, although a certain degree of mutual information is expected with means extracted from the bins. Second, an ideal-case variability study would focus on the entirety of the frequency (or period) space as a feature. However, it is limited by practical issues, such as the curse of dimensionality. Using the compressed periodogram as a basis for features allows to scale the pipeline to work with millions of light curves. We note that using the full periodogram as a feature is an interesting area of research to explore for smaller-scale classification problems.
    
    Both individual and averaged bin edges can reflect the underlying physics of hybrid pulsators. We observed finer g-mode bins due to rotational modulation, as rotational peaks can contribute significantly to the cumulative power distribution which does not discriminate between the sources of power. In stars with no rotational modulation, g-mode regime was typically binned sparser. Including a significant number of stars with rotational peaks in the training set results in finer low-frequency binning, especially with power law scaling.

\subsection{PU bagging classifier}

    Given that positively labelled instances (known hybrids) have been selected based on their attributes, our learning problem builds on the SAR — Selected at Random — assumption. This limits the available pool of PU Learning algorithms that can be applied. SAR-based classifiers do not incorporate priors, meaning that we do not inject the expected occurrence rate of hybrids into the model. In our framework, we used a PU bagging classifier \citep{mordelet2014}. Bagging is an example of ensemble learning, which involves using multiple weak learners — classifiers making individual classifications — and combining their results. Bagging specifically involves learning $T_{weak}$ weak learners from $i_{boot}$ bootstrapped samples of the entire dataset. 

    Due to the large size of the unlabelled set, we first randomly sampled 15\% of observations from each sector, to not introduce potential sector-specific biases. We then trained an ensemble of $T_{weak}=4000$ decision trees (weak learners), on a bootstrapped sample $i_{boot}$ of unlabelled instances equal to the size of the labelled sample, separately for nominal and extended TESS missions, and separately only using frequency- and period-domain features. Each of the trees makes an out-of-bag prediction — prediction on objects not in $i_{boot}$ — resulting in a distribution of the probability of being a hybrid for every unlabelled example. We then used the trained model to predict the probability of a hybrid nature for all observations from each sector.
    
    The final probability for each unlabelled target being a hybrid pulsator was determined as an average of $n_{oob}$ out-of-bag predictions associated with it. Notably, due to the all-sky nature of this study, multiple predictions were made for the same star in all TESS sectors it was available. We follow the `single sector - single instance' approach used by \citet{hey2024}, meaning that a star had multiple opportunities to be labelled a hybrid. We did not combine the predictions from different sectors to make a final prediction for a particular object, to avoid decreasing probabilities for genuine hybrids due to instrumental noise characteristics occurring in a single sector. Our final probability for a single light curve was an average of two predictions made separately using frequency- and period-domain features.

    An important note to make is that a normal grid search or cross-validation is not available for hyperparameter tuning in the PU Learning setting as there is no precision and therefore no f-score (mean of precision and recall). An alternative f-metric exists for some PU Learning methods, but only for the ones following the SCAR — Selected Completely at Random — assumption. On the other hand, our architecture has an advantage that it is based on decision trees, which are computationally inexpensive to learn. We therefore chose $T_{weak}=4000$, resulting in reduced uncertainty due to a high number of predictions for each unlabelled instance.

\subsection{Classification evaluation \& error analysis}

Evaluation of this classification framework is challenging as precision is not readily available, and consequently, nor is the f-score metric. While precision can be estimated \citep{claesen2015}, it is not exact as for binary classification and is therefore less reliable. We instead analysed our classification by inspecting high-probability predictions of some of the targets from our sample, making predictions for labelled non-hybrids from \citet{hey2024}, and analysing feature importance. 

\subsubsection{Recall}

Recall — the fraction of true positive examples recovered from the unlabelled set — is the only traditional evaluation metric available for SAR-based PU Learning methods. As PU Learning relies on an idea that some unlabelled positive instances exist in the unlabelled set (Fig.~\ref{fig:pu}), recall allows us to understand how good the model is at finding new genuine hybrids. Not knowing all positive instances contained in the unlabelled dataset, however, means that this metric is not as exact as for the binary classification but rather a good proxy for it. To estimate it, we set aside a 15\% of the labelled hybrids to be used as a test set.

\begin{figure}
    \centering
    \resizebox{\hsize}{!}{\includegraphics{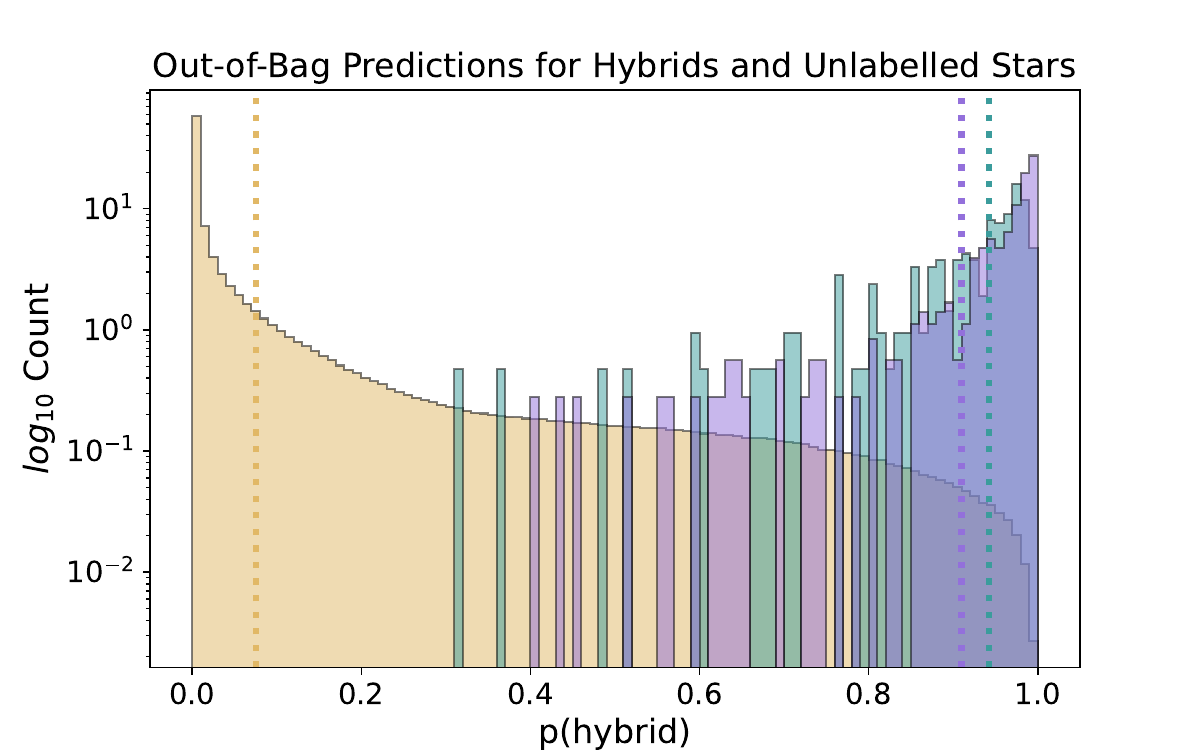}}
    \caption{Combined out-of-bag predictions for the unlabelled instances from nominal and extended missions with 15\% of the total data (golden) and hidden hybrids from nominal (teal) and extended (purple) missions. Teal (94.25\%) and purple (90.98\%) dotted lines represent mean predictions for hidden hybrids from nominal and extended missions, respectively. Golden dotted line represents the grand mean for all unlabelled targets (7.57\%) used for training.}
    \label{fig:recall}
\end{figure}

Our classifier reported a high recall, demonstrated in Fig.~\ref{fig:recall}, both in the nominal and extended TESS missions. The average out-of-bag prediction for all known hybrids was 93.04\%. 92.08\% receive average probabilities above 80\%, 88.56\% above 85\%, 79.75\% above 90\%, and 61.62\% above 95\%. Some hidden hybrids from the nominal mission received lower scores than extended mission hybrids. We assume that this is due to the larger volume of the unlabelled set, which generally requires more weak learners for a more reliable probability estimation. From the entire unlabelled subset used in training, only 1.00\% received probabilities above  80\%, 0.61\% above  85\%, 0.31\% above 90\%, and 0.09\% above 95\%. This suggests that hybrids construe only a small fraction of dwarf stars; we revisit this in Section \ref{sec:catalog}.

\subsubsection{Estimating contamination from rotational modulation and pure g- and p-mode pulsators}

\begin{figure}
    \centering
    \resizebox{\hsize}{!}{\includegraphics{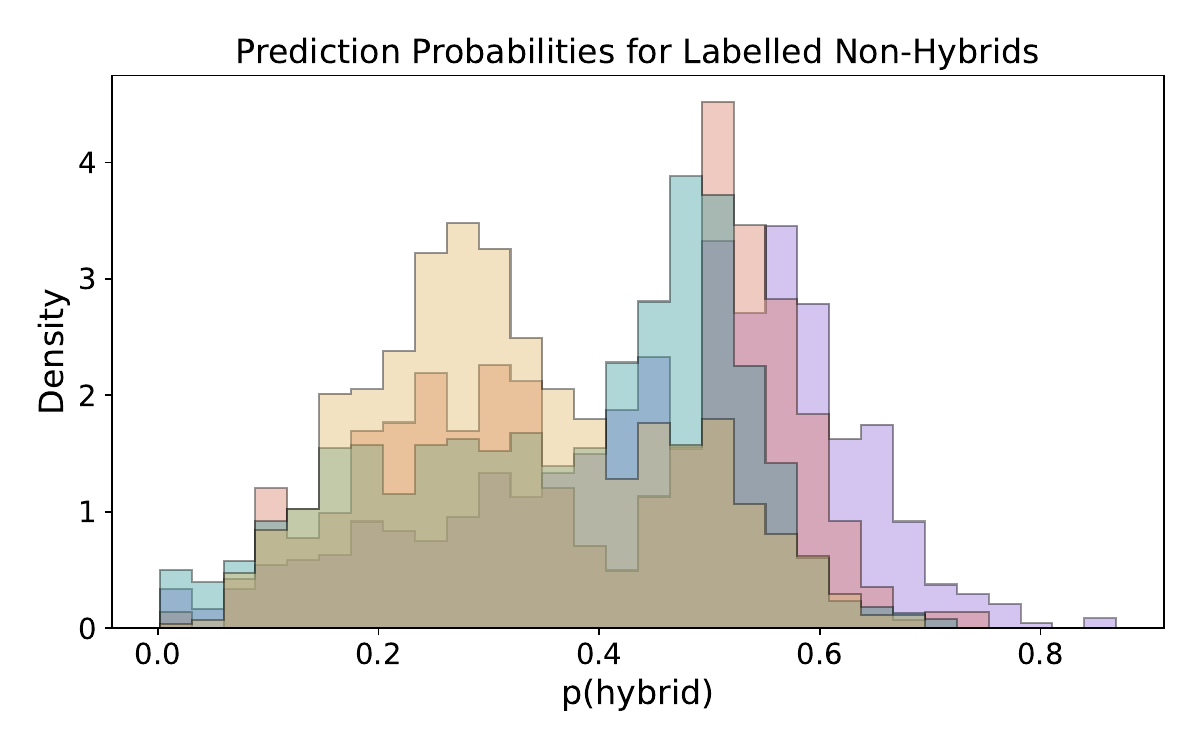}}
    \caption{Out-of-bag predictions for pure $\delta$ Sct stars (purple), eclipsing binaries (burgundy), pure $\gamma$ Dor / SPB pulsators (teal), and rotational variables (golden) from TESS sectors 1 and 14 from \citet{hey2024}.}
    \label{fig:hey}
\end{figure}

Distinguishing $\gamma$ Dor / $\delta$ Sct hybrids in a vast unlabelled sample is especially challenging due to the presence of other types of pulsators and, most critically, their parent classes: pure $\gamma$ Dor and $\delta$ Sct variables. Another potential source of confusion is the class of rotational variables, which is known to be cross-contaminated with g-mode pulsators from supervised variable star classification results by \citet{audenaert2021} and \citet{hey2024}. Although the unresolved comb of frequencies of g-mode pulsators is relatively easy to differentiate manually from one or three rotational peaks, it is not a trivial task for an automated method, especially without harmonic analysis and prewhitening (\citeauthor{debosscher2007} \citeyear{debosscher2007}, \citeauthor{richards2011} \citeyear{richards2011}) which we explicitly did not use in this study due to the volume of the unlabelled dataset and undesired additional data manipulation. In our framework, this issue leads to the expectation that some high probabilities may be assigned to $\delta$ Sct stars with rotational modulation instead of being a hybrid pulsator. To test that, we predicted the hybrid class probability (Fig.~\ref{fig:hey}) on a subset of the three potentially confusing classes: $\gamma$ Dor, $\delta$ Sct, and rotational variables, from \citet{hey2024}. We also predicted probabilities for a class of eclipsing binaries, as there could potentially be a (hybrid) pulsation component in a binary system, whose orbital period is found in the realm of (sub-)multiples of g~modes.

As expected, the class containing $\delta$ Sct stars received the highest out-of-bag scores, with a mean of 0.45. Pure g-mode pulsators have a mean score of 0.37. We note that this class includes both $\gamma$ Dor and SPB stars that can not be differentiated from each other without $T_{\mathrm{eff}}$ (\citeauthor{audenaert2021} \citeyear{audenaert2021}, \citeauthor{aerts2023} \citeyear{aerts2023}), so predicting on it with our classifier is essentially the same as predicting on pure $\gamma$ Dor stars. Eclipsing binaries and rotational variables received mean scores of 0.39 and 0.32, respectively. This allows us to conclude that our classifier should in principle be able to differentiate genuine hybrids from variable stars with variability in just one of the frequency regimes.

Given the distribution of dominant variability of the labelled set (Fig.~\ref{fig:A1}), we inspected the highest probability predictions for this non-hybrid dataset. The highest probabilities have all been assigned to the $\delta$ Sct class, but only three sources receive scores above 80\%. This suggests that placing a probabilistic threshold at 80\% would allow one to avoid most of the contamination coming from variable but non-hybrid sources. For the remainder of this paper we thus used the 80\% cut-off for our candidate targets.

\subsubsection{Feature importance}

Feature importance allow us to understand how much each feature contributes to the model’s predictions. While it does not directly evaluate classification performance, it provides a valuable insight into whether the model is working as intended. The importances of features of our models shared certain similarities. Inspecting individual feature importances revealed that the most important features for frequency-based models were associated with higher-frequency bins responsible for capturing p-mode variability. The importance of features of period-based models were the highest for short-period bins, also capturing p-mode variability. This can be seen as a result of the predominance of hybrids where $f_{1}$ is a p~mode in the training set (Fig.~\ref{fig:A1}).

\begin{figure}
    \centering
    \resizebox{\hsize}{!}{\includegraphics{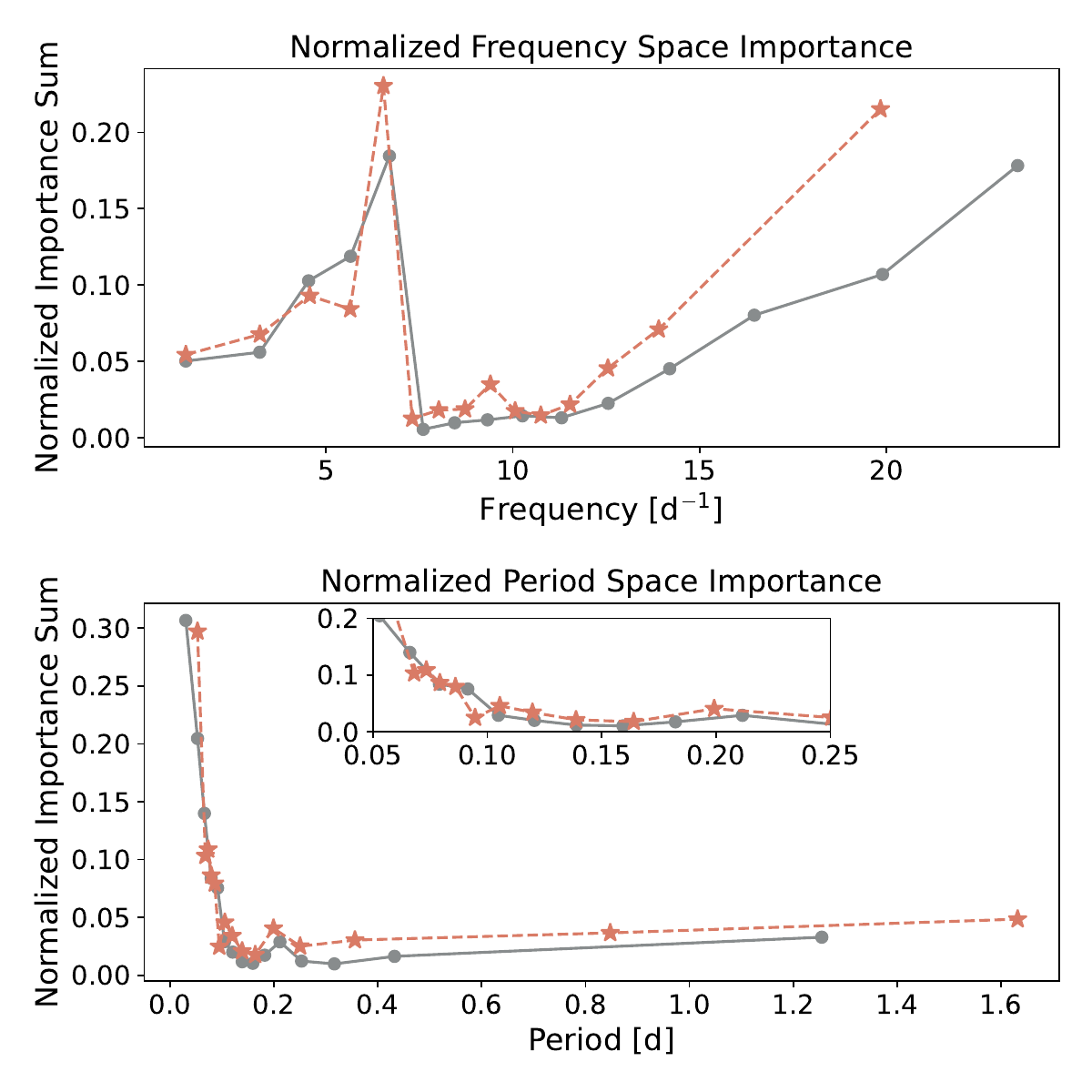}}
    \caption{Normalized importance of the frequency (top) and period (bottom) space, plotted at the centre of each frequency bin of nominal (burgundy) and extended (grey) mission models. Note that the final frequency bin goes to the Nyquist limit and the final period bin goes to approximately the inverse of the closing edge of the first frequency bin — they are plotted in their respective places for visibility.}
    \label{fig:top_bins}
\end{figure}

Summing the importances of the five features associated with each of the fifteen frequency bins (Fig.~\ref{fig:top_bins}) confirmed that both models favoured high-frequency regimes. For frequency-based classifiers, the low-frequency regime was captured by the first three bins, which had moderately high importance ($>0.05$). High importance was attributed to the first two high-frequency bins. There is an increasing importance trend in the frequency space above 10 d$^{-1}$, where one might expect genuine p~modes or their aliases mirrored around the Nyquist limit \citep{hey2021}, which results from the cumulative power distribution summing the power regardless of its origin. The importance of bins of the period-based classifiers also had a high preference for p-mode variability similar to their frequency counterparts. 

From the feature importance and the summed importance of the bins, we could expect both a predominance of p-mode dominated hybrids in the high-probability candidates from the unlabelled set and p-mode contaminants. However, combining the prediction of two classifiers for a sample is expected to alleviate some of this bias, producing a more balanced classification result.

\section{All-sky candidate hybrids catalogue}

\subsection{Frequency analysis}\label{sec:catalog}

After classification, we filtered the data to remove potential contamination by false positives and contamination by instrumental effects. From a Lomb-Scargle periodogram for high-probability candidate hybrids $(p(hybrid) > 0.8)$ we filtered out those that did not have at least one significant peak in frequency regimes >4 d$^{-1}$ ($f_{g}$) and <4 d$^{-1}$ ($f_{p}$) with a <1\% false-alarm probability (FAP). For this filter, $f_{p}$ was the highest peak that was not an integer or half-integer multiple in the range of $[1.5, 15.5]$ of $f_{g}$ to also avoid potential contamination by pure g-mode pulsators with harmonics in higher frequencies and eclipsing binaries with no pulsation component. We also removed candidates where a signal with >1\% FAP check was a near-zero frequency (mainly for the nominal mission), 0.9, or 1 d$^{-1}$ (mainly for the extended mission), within twice the TESS sampling frequency $f_{sample}$. These three filters left us with 154,778 out of 284,325 light curves. The final sample was further reduced to 62,026 by applying a $T_{\mathrm{eff}} \geq 6,000$ cut from \gaia DR3 for homogeneity with the training set (see Section \ref{sec:params}). This resulted in a final fraction of candidate hybrids of 1.9\% and 2.2\% of the total analysed targets in nominal and extended mission datasets, respectively.

\begin{figure*}
    \centering
    \includegraphics[width=12cm]{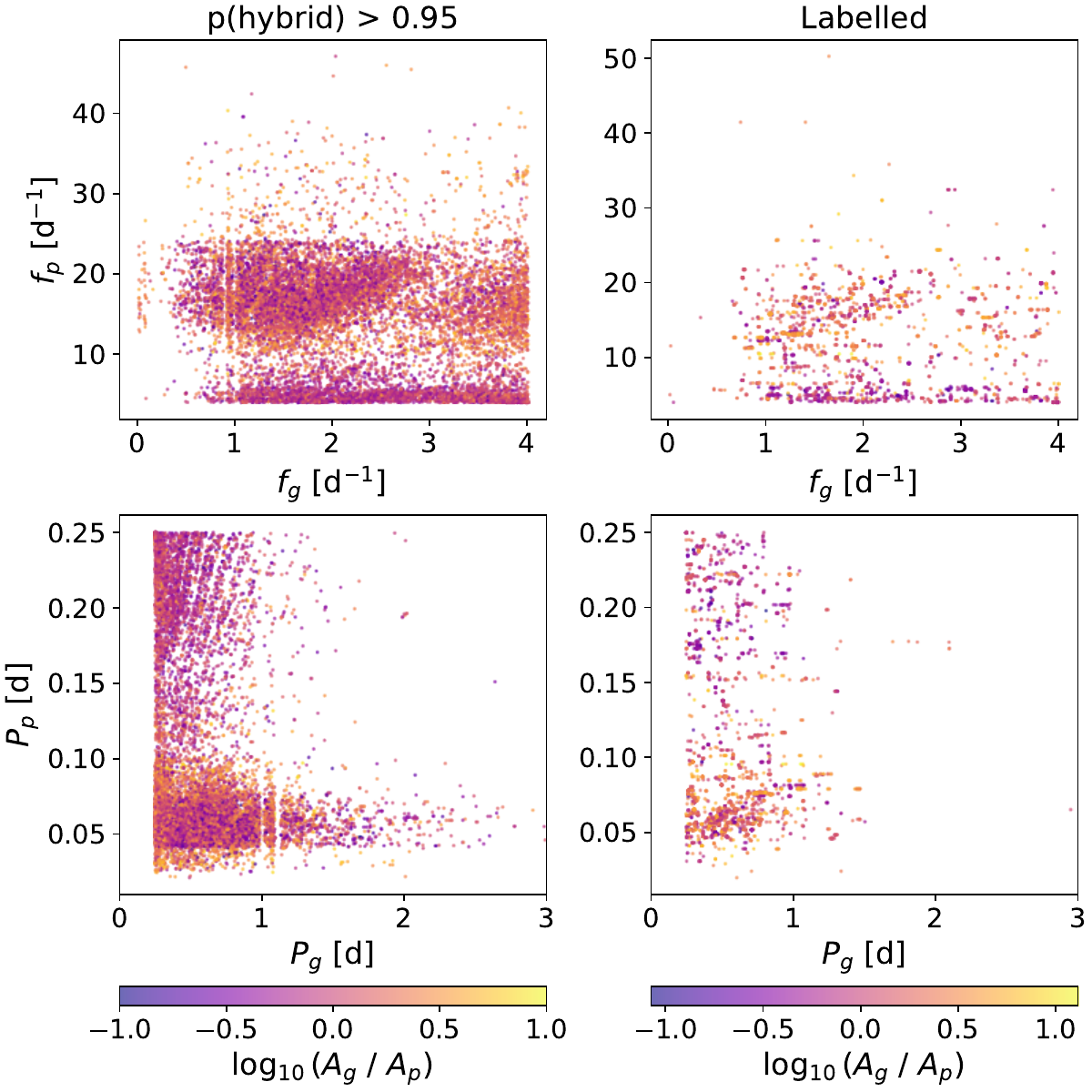}
    \caption{Frequency-frequency (top) and period-period (bottom) plot of candidate hybrids with $p(hybrid) > 0.95$ (left) and labelled hybrids (right) for highest peak detected in frequency regimes <4 d$^{-1}$ ($f_{g}$) and >4 d$^{-1}$ ($f_{p}$) with a <1\% FAP. The colour bar represents $log_{10}$ amplitude ratios of the two frequencies (periods).}
    \label{fig:populations}
\end{figure*}

In Fig.~\ref{fig:populations} we show the frequencies associated with the highest frequencies in the <4 d$^{-1}$ and >4 d$^{-1}$ frequency (top left) and corresponding period (bottom left) regimes with only candidates, with only $p(hybrid) > 0.95$ plotted for visibility. The frequency-frequency relation (top left) shows a multi-modal distribution with respect to both $f_{g}$ and $f_{p}$, potentially linked to differences in internal structure of these stars. In the period space (bottom left), we found two distinct groups of pulsators with short — a dense population with $P_g \lesssim 1$ with a wide range of $P_p$ — and long-period — a less numerous subset with $P_g \gtrsim 1$ and typically lower $P_p$ values — g~modes, potentially indicative of differences in the spherical degree $l$ of their dominant pulsation modes (Chapter 7 of \citeauthor{aerts2010} \citeyear{aerts2010}). As future work, we intend to investigate period spacing patterns of targets in these two clusters and compare them with stars with known identified modes \citep{li2020}. We also note that a number of hybrids in the training set are fast rotators and some of the manually inspected high-probability candidates show rotational splittings. This makes these stars suitable candidates for further studies to determine internal rotational properties and angular momentum transport mechanisms in AF stars \citep{aerts2025}.

The harmonic avoidance criterion applied together with the FAP filter is responsible for a large portion of the targets cut. We show the harmonic behaviour of candidates without such a filter on Fig.~\ref{fig:A3-1}, which depicts a period-period relation where $P_{p}$ is simply the inverse of the highest peak in the >4 d$^{-1}$ frequency regime. The targets with low log-amplitude ratios are sources where an extracted $P_{p}$ was a harmonic of $P_{g}$ instead of a genuine p~mode. We manually inspected sources from $P_{p} = 2 \times P_{g}$, $P_{p} = 3 \times P_{g}$, $P_{p} = 4 \times P_{g}$, and $P_{p} = 5 \times P_{g}$ (one example of each is shown on Fig.~\ref{fig:A3-2}). Some were eclipsing binaries, potentially with a pulsational component. This is because a feature discriminative of this type of stars is in the time domain \citep{hey2024}, which we did not consider in this study. We note that increasing the probabilistic threshold (e.g. to $p(hybrid) > 0.95$ on Fig.~\ref{fig:A3-1}, right) filtered these sources out, meaning they receive lower probabilities than genuine hybrids. This suggests that the initial probabilistic cut at $p(hybrid) > 0.8$ might be too generous, depending on the nature of the foreseen asteroseismic follow-up study.

\begin{figure}
    \centering
    \resizebox{\hsize}{!}{\includegraphics{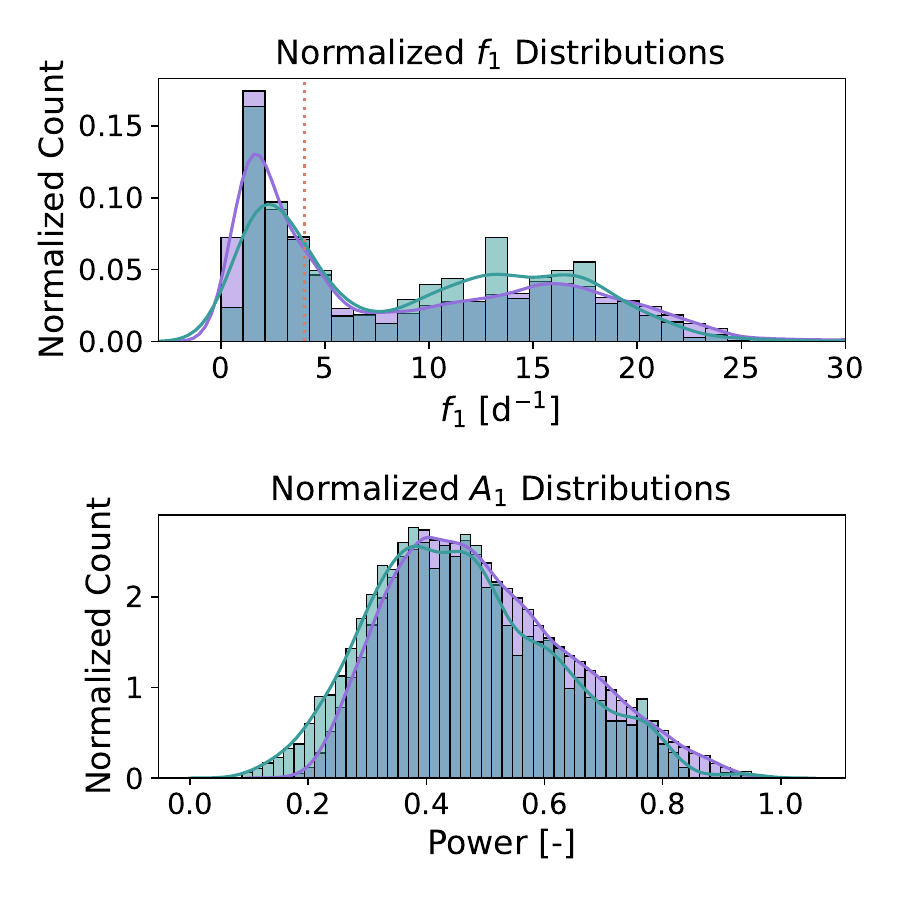}}
    \caption{Densities of $f_{1}$ (top) and $A_{1}$ (bottom) for hybrids in the labelled dataset (teal) and high-probability candidates (purple). Histograms are plotted from 1,000 samples for each candidate within the uncertainty range and a Kernel Density Estimator is plotted directly from point estimates of $f_{1}$ and $A_{1}$. A burgundy dotted line represents a 4 d$^{-1}$ frequency dividing the bimodal distribution into the g- and p-mode regimes.}
    \label{fig:f1}
\end{figure}

As can be seen on Fig.~\ref{fig:f1}, most labelled hybrids are p-mode dominated, and so are the detected candidates. The distribution of power in the candidate catalogue is partly linked to biases propagated through the model. Our model, however, did not worsen this bias, evidenced by the distribution (Fig.~\ref{fig:f1}, top). The shapes of distributions for both $f_{1}$ of two populations are not identical but resemble each other, and so are the ones for $A_{1}$. The distribution of $A_{1}$ for the unlabelled set is shifted to the right, suggesting that we found higher-amplitude targets than we have in the labelled set. 

\begin{figure}
    \centering
    \resizebox{\hsize}{!}{\includegraphics{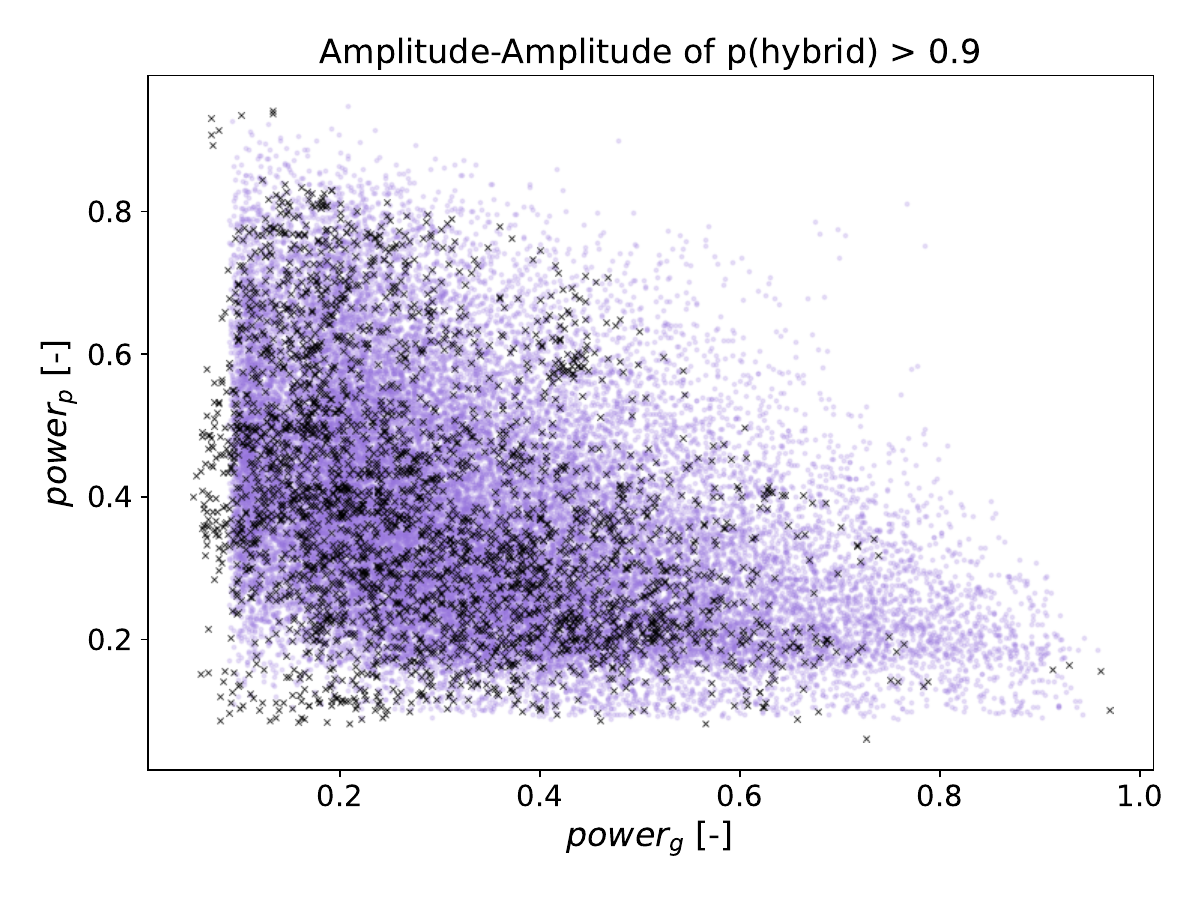}}
    \caption{Amplitude-amplitude plot of the highest peak detected in frequency regimes <4 d$^{-1}$ ($A_{g}$) and >4 d$^{-1}$ ($A_{p}$) with a <1\% FAP of candidate hybrids with $p(hybrid) > 0.9$ (purple) and labelled hybrids (black).}
    \label{fig:amps}
\end{figure}

This is confirmed by Fig.~\ref{fig:amps} (only sources with $p(hybrid) > 0.9$ are plotted for visibility), which shows that there are very few candidates with low amplitudes in both p- and g-mode regimes (cluster on the bottom left), while more labelled hybrids were found in the same region. It also shows that more hybrids cluster to the left, meaning that the detected population is dominated by oscillators with a lower-amplitude g-mode component. Additionally, we found more hybrid candidates with a strong component in one of the regimes and a weaker one in the other than stars with equally strong p and g~modes, as was also the case for the training set examples.

The density of high-amplitude p~modes is higher than the one for g~modes — further confirming that most candidates are p-mode dominated. There are several explanations for this. First, single-sector TESS observations do not have sufficient frequency resolution to fully resolve g~modes, as concluded in Section \ref{sec:kepler}. One potential way to remedy this would be to stitch together observations from multiple sectors and classify longer light curves. Second, pure $\delta$ Sct stars are somewhat rare in space photometry \citep{grigahcene2010} as they have could have a g~mode if the quality of the light curve is high enough, so a p-mode pulsator is more likely to be labelled a hybrid. Finally, the features used in this study relied on the cumulative power distribution, which does not discriminate against the origin of the variability, be that a genuine pulsation, a rotational peak, or a systematic at low frequencies. This gave a p-mode pulsator more chances to be confused with a hybrid than a g-mode pulsator that already has peaks in the same frequency regime.

Following the approach of \citet{hey2024}, we performed two-sample Kolmogorov–Smirnov tests with 95\% confidence level for the distributions of amplitudes and frequencies of the dominant modes on labelled hybrids, the final list of high-probability candidates following our three filtering stages, and pure p- and g-mode pulsators from \citet{hey2024}. The null hypothesis was that the two distributions are sampled from the same distribution. Each of our comparisons reported a p value around $10^{-4}$. For the comparison of labelled hybrids and high-probability candidates, this reinforced the underlying SAR assumption of PU Learning, suggesting that the labelled hybrids had not been chosen randomly from the total population of hybrids but rather for their specific attributes. When comparing both labelled hybrids and high-probability candidates with pure p- and g-mode pulsators, this can be understood in terms of several excitation mechanisms contributing to the oscillations of hybrids (\citeauthor{hey2024} \citeyear{hey2024}, \citeauthor{mombarg2024} \citeyear{mombarg2024}). Further exploitation of the sample could focus on comparing the distributions of fundamental parameters or pulsation properties beyond $f_{1}$ and $A_{1}$.

\subsection{Fundamental parameters}\label{sec:params}

For our final candidate selection, we investigated the fundamental parameters, namely effective temperature $T_{\mathrm{eff}}$, stellar luminosity $L$, and radius $R$. We first computed $L$  following the method from \citet{fritzewski2024} by calculating extinction-corrected absolute magnitudes and applying bolometric corrections to get the bolometric magnitude using \gaia\footnote{https://gea.esac.esa.int/archive/} photometry and GSP-Phot parameters. We then computed the radius from the Stefan-Boltzmann Law. We estimated the uncertainties of these parameters using standard uncertainty propagation rules from \gaia DR3 properties and assuming that $\sigma_{BC}=0.05$. As stated in Section \ref{sec:catalog}, we initially found that the distribution of $T_{\mathrm{eff}}$ for detected candidates extended further to cooler stars than for the labelled set (similar to what was found in \citeauthor{deridder2023} \citeyear{deridder2023}), which is a result of us including stars with $T_{\mathrm{eff}} \geq 4,000$ K in the unlabelled set and TIC and \gaia DR3 reporting different temperatures. In this study, high-probability candidates with $6,500 \geq T_{\mathrm{eff}} \geq 4,000$ were later removed for homogeneity with training hybrids.

\begin{figure*}
    \centering
    \includegraphics[width=17cm]{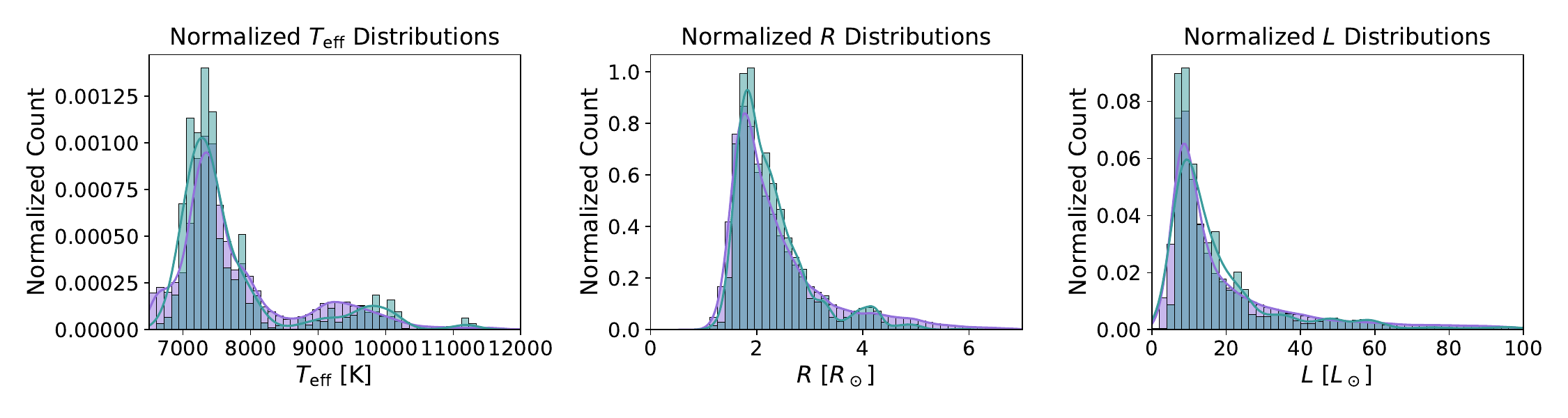}
    \caption{Distribution of fundamental parameters — $T_{\mathrm{eff}}$ (left), $R$ (middle), and $L$ (right) — of the final filtered sample of high-probability candidates (purple) and labelled hybrids (teal). Histograms are plotted from 1,000 samples for each candidate within the uncertainty range and a Kernel Density Estimator is plotted directly from point estimates of $T_{\mathrm{eff}}$, $R$, and $L$.}
    \label{fig:params}
\end{figure*}

We also observed (Fig.~\ref{fig:params}, left) that a number of candidates showed $T_{\mathrm{eff}} \geq 10,000$ K. This is in line with 
\citet{deridder2023},
\citet{aerts2023}, \cite{hey2024}, and \citet{mombarg2024}, who have already found that the transition from $\gamma$ Dor / $\delta$ Sct to SPB / $\beta$ Cep pulsators is not a sharp cut-off but rather a continuum, in contrast to predictions from instability computations. These studies also already reminded us that the impact of fast rotation is to make the stars look seemingly cooler in the 
Hertzsprung-Russel diagram \citep{deridder2023}. Given that our classifier had no information on the temperature of targets, it is therefore not surprising they are found in the high-probability dataset. Considering their significant asteroseismic potential \citep{pedersen2021}, we opted not to remove them from the candidate list. We finally note that the distribution of radii and luminosity for both labelled hybrids and the high-probability candidates follow each other closely.

We prepared a publicly available catalogue of high-probability candidates ($p(hybrid) > 0.8$) with TIC and \gaia DR3 identifiers, coordinates, TESS sector, highest frequencies in the low ($f_{g}$) and high ($f_{p}$) frequency regimes, their amplitudes ($A_{g}$ and $A_{p}$, respectively), global properties ($T_{\mathrm{eff}}$, $L$, $R$), and probabilities of being considered a hybrid (Table 1). Examples of light curves are provided in Fig.~\ref{fig:A4}. From our catalogue, there is an overlap of 1,619 individual sources labelled as hybrid and 464 as other classes, mainly p-mode pulsators, from the nominal mission catalogue by \citet{hey2024} and an overlap of 21 pulsating eclipsing binaries from \citeauthor{ijspeert2021} \citeyear{ijspeert2021} and Kemp et al. (subm.), of which 9 are labelled as g-mode, 10 p-mode, and 2 hybrid pulsating binaries. We additionally prepared high-probability candidates that passed the $T_{\mathrm{eff}}$ and $f_{g}$ filtering but not the FAP threshold in one of the frequency regimes as respective catalogues of candidate p- (Table 2) and g-mode (Table 3) pulsators.

\section{Conclusion}

In this paper, we analysed TESS light curves of known $\gamma$ Dor / $\delta$ Sct hybrid pulsators from datasets of confirmed hybrids and a large dataset of stars using TGLC light curves, resulting in a substantial increase in the available target pool of candidate hybrids. A comparison of TGLC light curves and their dominant and secondary periodic variability extracted from them generally agrees very well with other available light curves. TGLC light curves are at least as suitable as QLP light curves for hybrid pulsator studies. However, we note that the former have a magnitude limit that potentially allows to increase the current pool of available sources: 19.89\% of our candidates are fainter than the magnitude limit of QLP. The two pipelines are both equally capable of recovering dominant \kepler-extracted frequencies for the labelled targets. Both are, however, better at detecting high-frequency signal than low frequencies, as single-sector TESS observations do not have enough frequency resolution to detect g-mode variability in an optimal way.

We performed a semi-supervised classification of hybrid pulsators on 29 million TGLC light curves with a PU Learning classifier.\footnote{The code used in the framework of this study is available at https://github.com/nikkliapets/smartbinning\_pulsators.} This classification is a challenging learning problem as it essentially combines two different goals — differentiating pulsating stars from a largely non-variable set and a specific class from other variable stars, including its parent classes. This is normally achieved by relying on different features. Our method additionally overcomes the need to have a representative labelled set in traditional supervised machine learning architectures. Our study demonstrated that by leveraging a small dataset of known rare objects, we are able to find many new candidates by using features specifically tailored to a class of interest and a machine learning method suitable for learning problems under class imbalance. Such projects of looking for objects of a particular rare class can be seen as the first necessary step for population studies. The volume of data analysed in this project also demonstrated the practical aspects of employing automated pipelines in the age of space missions.
    
Our classification resulted in a catalogue of over 62,000 new candidate main-sequence hybrid pulsator light curves with significant peaks in both high- and low-frequency regimes (with additional catalogues of pure p- and g-mode pulsator candidates). These are ideal starting points for many asteroseismic follow-up studies. While our hybrid catalogue is dominated by hybrid candidates with higher p~modes, it is an expected result of the TESS sampling rates, the composition of our labelled training set, and the occasional low-frequency instrumental systematics.

The population of detected hybrids comprises various sub-populations with distinct pulsation properties. Future exploitation of these stars can focus on the following:

    \begin{itemize}
       \item From a machine learning perspective: unsupervised clustering of the sample to reveal different sub-classes of hybrids, i.e. $\gamma$ Dor / $\delta$ Sct hybrid, $\delta$ Sct / $\gamma$ Dor hybrids, hybrids with rotational and magnetic splittings, hybrids with rotational modulation, etc.; and

       \item From an astrophysical perspective: exploitation of internal physical properties utilizing information provided by both p and g~modes, particularly to measure differential rotation, mode coupling, and more precise measurements of asteroseismic ages and other astrophysical parameters from asteroseismology (\citeauthor{aerts2021} \citeyear{aerts2021}, \citeauthor{aerts2024} \citeyear{aerts2024}).
   \end{itemize}

Other types of hybrid pulsators (SPB / $\beta$ Cep hybrids and p- and g-mode pulsating sdBVs) were not included in this study. However, our methodology is also applicable to these stars too. Moreover, we found B-type hybrids, as demonstrated by the distribution of $T_{\mathrm{eff}}$ of detected candidates. In this context, it is interesting to look into developing machine learning architectures that can differentiate different classes of hybrids (e.g. $\gamma$ Dor / $\delta$ Sct hybrids from SPB / $\beta$ Ceph hybrids) by utilizing observables from other surveys, such as \gaia data \citep{brown2016} or including other modalities, i.e. spectroscopy, such as \gaia \citep{huijse2025} or SDSS-V \citep{kollmeier2019} spectra. Furthermore, additional efforts can be put into more precisely determining the probabilistic threshold of what is considered a hybrid, which would allow us to put a prior on the occurrence of these stars, opening avenues for using semi-supervised learning methods build on the SCAR assumption, which are normally easier to evaluate \citep{bekker2020}.

Finally, with the upcoming PLATO \citep{rauer2025} mission that will observe brighter stars with at least two year-long baseline, methods specifically tailored to the study of hybrids with dominant g~modes are of particular interest to the variable star community. We stress that 4,283 of our high-probability candidates fall in PLATO's first field of view (\citeauthor{nascimbeni2025} \citeyear{nascimbeni2025}, \citeauthor{jannsen2025} \citeyear{jannsen2025}), offering an opportunity for their asteroseismic exploitation to prepare for future PLATO data releases.

\section{Data availability}

Tables 1, 2, and 3 are only available electronically at the CDS: https://cdsarc.cds.unistra.fr/viz-bin/cat/J/A+A/703/A240.

\begin{acknowledgements} 
    MK and CA acknowledge The Kavli Foundation for their financial support in the framework of the Kavli Scholarship given to MK. CA and PH acknowledge financial support from the European Research Council (ERC) under the Horizon Europe programme (Synergy Grant agreement N°101071505: 4D-STAR). While partially funded by the European Union, views and opinions expressed are however those of the author(s) only and do not necessarily reflect those of the European Union or the European Research Council. Neither the European Union nor the granting authority can be held responsible for them. AK, DJF, and CA acknowledge financial support from the Flemish Government under the long-term structural Methusalem funding programme by means of the project SOUL: Stellar evolution in full glory, grant METH/24/012 at KU Leuven. This paper includes data collected by the TESS mission. Funding for the TESS mission is provided by the NASA’s Science Mission Directorate. This work has made use of data from the European Space Agency (ESA) mission \gaia (https://www.cosmos.esa.int/gaia), processed by the \gaia Data Processing and Analysis Consortium (DPAC, https://www.cosmos.esa.int/web/gaia/dpac/consortium). Funding for the DPAC has been provided by national institutions, in particular the institutions participating in the \gaia Multilateral Agreement. 
\end{acknowledgements}

\bibliographystyle{aa}
\bibliography{bib.bib}

\begin{appendix}
\section{Frequency comparisons}

\begin{figure}[H]
    \centering
    \includegraphics[width=0.85\textwidth]{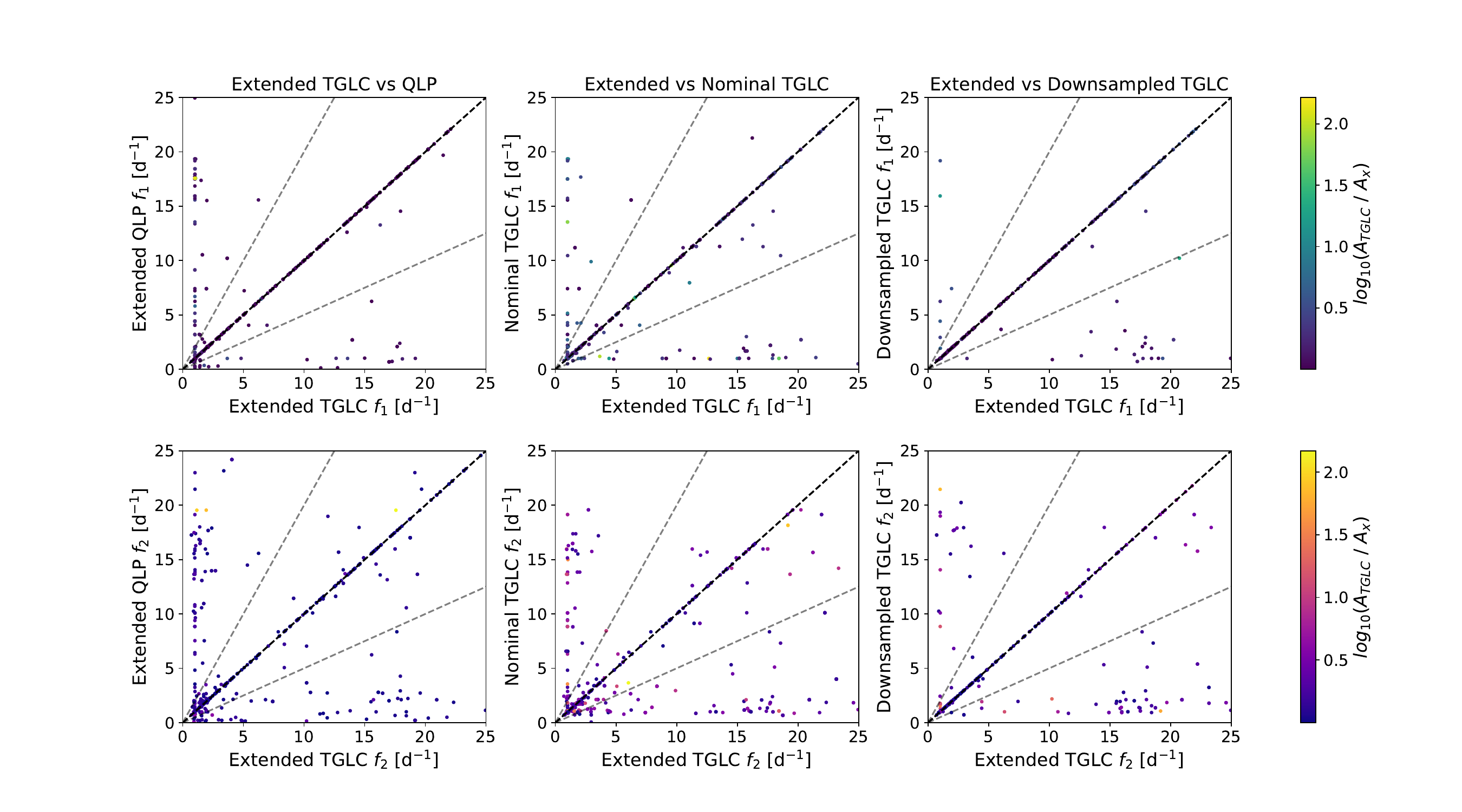}
    \caption{Comparison of $f_{1}$ (top) and $f_{2}$ (bottom) of confirmed hybrids in extended mission TGLC with extended mission QLP (left), nominal mission TGLC (middle), and extended mission TGLC downsampled to nominal mission quality (right). The black line indicates unity and two grey lines are double and half unity. The colour bar represents $log_{10}$ amplitude ratios of the two compared observations. Only sources with frequencies of up to 25 d$^{-1}$ are shown for visibility.}
    \label{fig:A1}
\end{figure}

\section{PU learning schematic}

\begin{figure}[H]
    \centering
    \includegraphics[width=8cm]{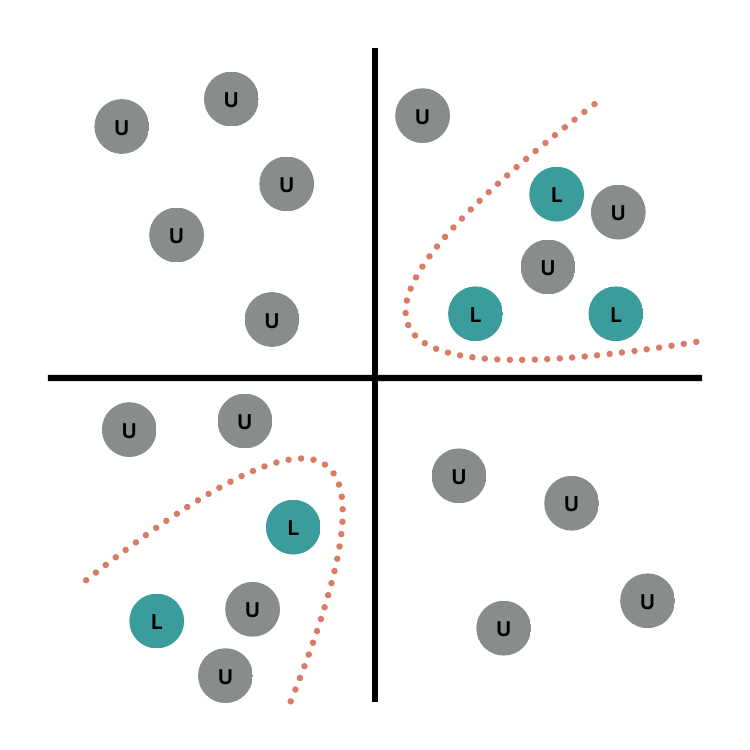}
    \caption{Graphic representation of a 2D projection of a PU Learning classifier (burgundy) looking for unobserved positive instances within a larger dataset of unlabelled instances (grey) with the help of a smaller labelled set (teal). Datasets not to scale.}
    \label{fig:pu}
\end{figure}

\clearpage

\section{Harmonic behaviour}

\begin{figure}[htbp]
    \centering
    \includegraphics[width=13cm]{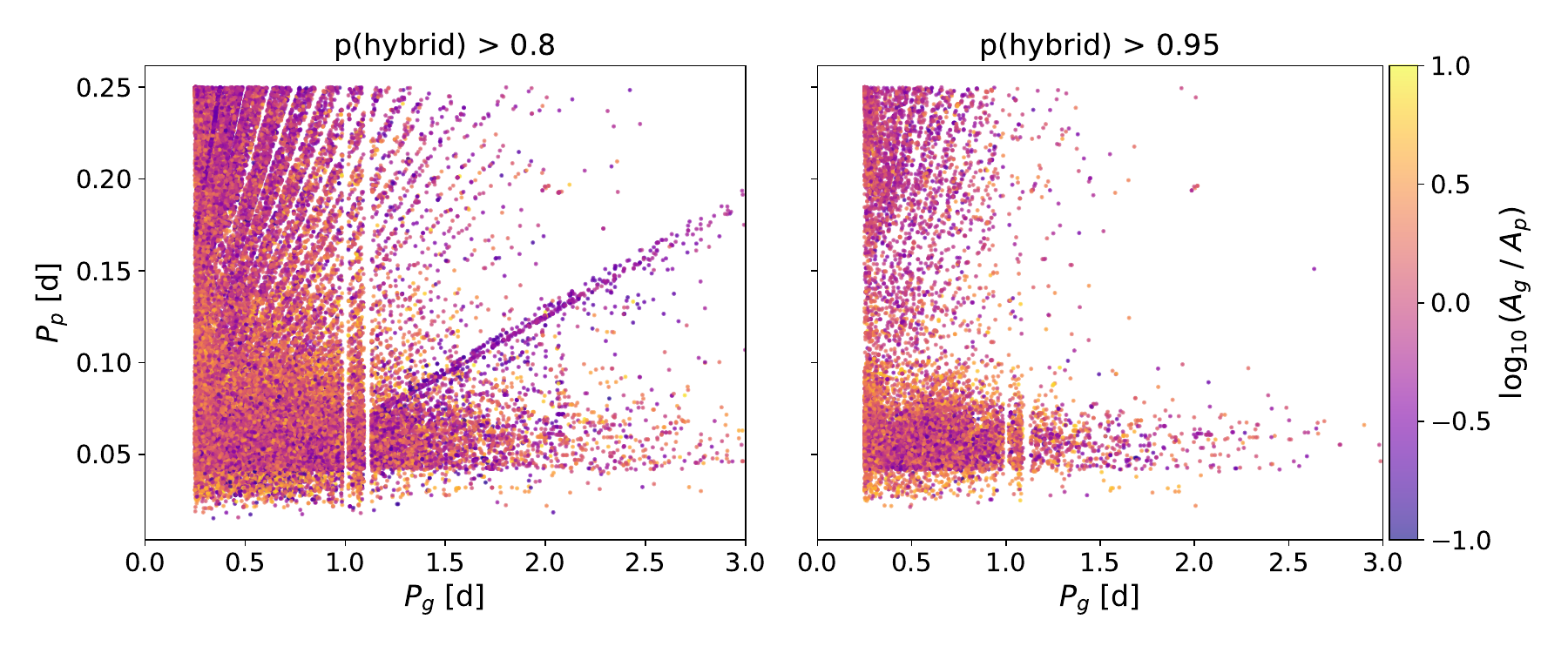}
    \caption{Period-period plot of candidate hybrids with $p(hybrid) > 0.8$ (left) and $p(hybrid) > 0.95$ (right) for the inverse of the highest peak detected in frequency regimes <4 d$^{-1}$ ($f_{g}$) and >4 d$^{-1}$ ($f_{p}$) with a <1\% FAP without harmonic avoidance criterion. Note the diagonal lines where $P_{p}$ is an integer multiple of $P_{g}$. The colour bar represents $log_{10}$ amplitude ratios of the two periods.}
    \label{fig:A3-1}
\end{figure}

\begin{figure}[htbp]
    \centering
    \includegraphics[width=13cm]{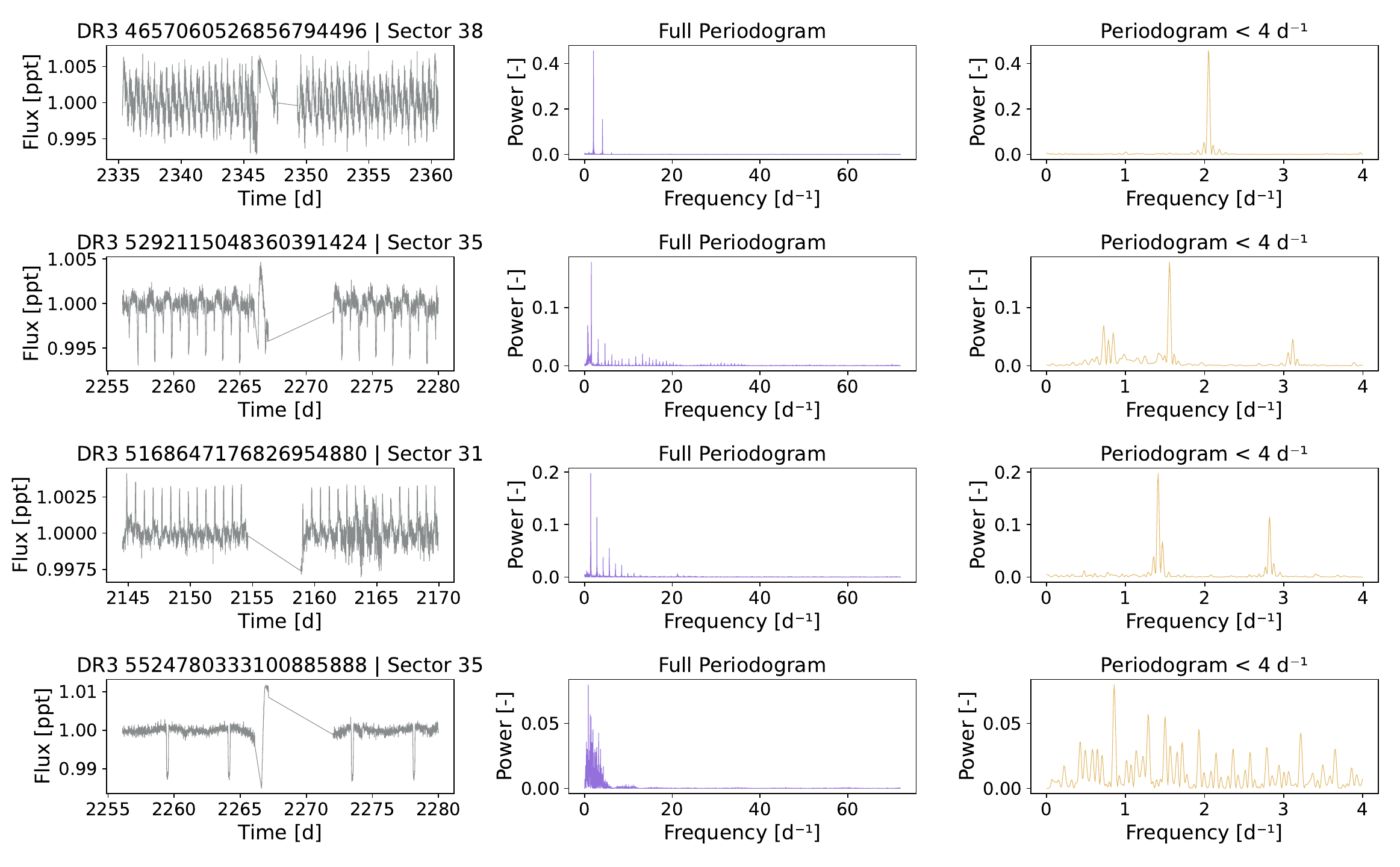}
    \caption{Light curve (left), periodogram (middle), and zoomed-in periodogram (right) of a single source where $P_{p} = 2 \times P_{g}$, $P_{p} = 3 \times P_{g}$, $P_{p} = 4 \times P_{g}$, and $P_{p} = 5 \times P_{g}$, respectively.}
    \label{fig:A3-2}
\end{figure}

\clearpage

\section{Examples of candidates}

\begin{figure}[htbp]
    \centering
    \includegraphics[width=13cm]{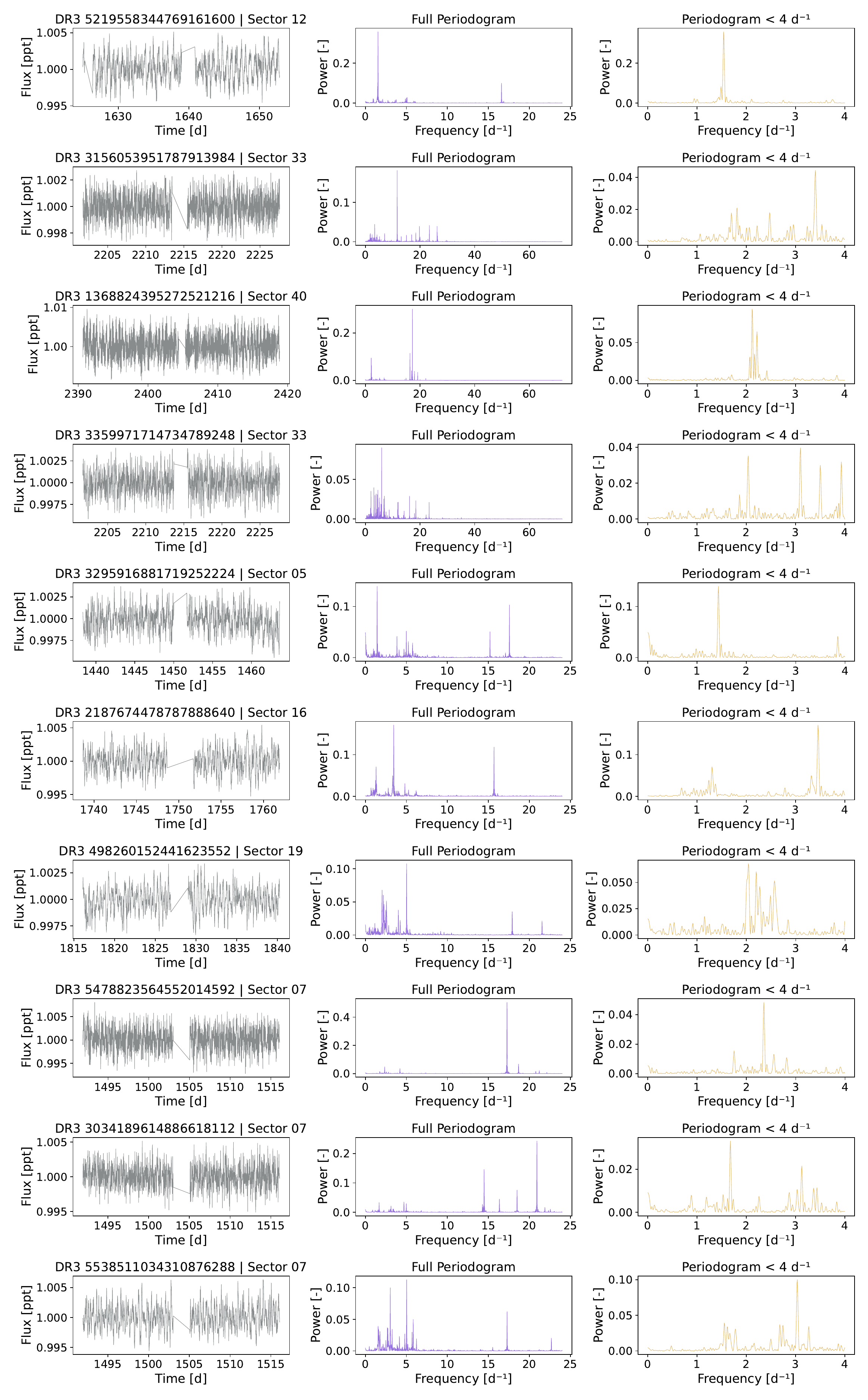}
    \caption{Light curve (left), periodogram (middle), and zoomed-in periodogram (right) of ten high-probability candidates of being a hybrid pulsator. Note the different Nyquist limit for nominal (rows 1, 5-10) and extended (rest) mission candidates.}
    \label{fig:A4}
\end{figure}

\end{appendix}

\end{document}